\documentclass{aa}  
\usepackage{graphicx}
\usepackage{adjustbox}
\usepackage{amsmath}
\usepackage{color}
\usepackage{txfonts}
\usepackage{hyperref}

\usepackage{amsmath}
\usepackage{xcolor}

\usepackage{media9}

\begin{document} 

\title{Origin of radio polarization in pulsar polar caps}

\titlerunning{Origin of radio polarization in pulsar polar caps}

\author{Jan Ben\'{a}\v{c}ek \inst{1,2},
      Axel~Jessner\inst{3}, 
      Martin Pohl \inst{1,4},
      Tatiana Rievajová \inst{5},
      and Lucy S. Oswald \inst{6}
      }
      
\authorrunning{Benáček et al.}

\institute{
    Institute for Physics and Astronomy, University of Potsdam, 14476 Potsdam, Germany \\ \email{jan.benacek@uni-potsdam.de}
    \and
    Astronomical Institute of the Czech Academy of Sciences, 25165 Ondřejov, Czech Republic
    \and
    Max-Planck Institute for Radio Astronomy, 53121 Bonn, Germany
    \and
    Deutsches Elektronen-Synchrotron DESY, Platanenallee 6, 15738 Zeuthen, Germany
    \and 
    Institute of Theoretical Physics and Astrophysics, Masaryk University, 611 37 Brno, The Czech Republic
    \and 
    School of Physics \& Astronomy, University of Southampton, Southampton SO17 1BJ, United Kingdom
}

\date{Received: ; accepted:}

  \abstract
   {
   It is crucial to know the polarization properties of coherent radio waves that escape from pulsar polar caps to calculate the radiative transfer through the magnetosphere and to predict observable radio properties. 
   }
   {
   We describe pair cascades in the pulsar polar cap, and we determine for the first time the Stokes parameters of the escaping radio waves from first-principle kinetic simulations for a pulsar with a magnetic obliquity of $60^{\circ}$.
   }
   {
   We present 3D particle-in-cell kinetic simulations that include quantum-electrodynamic pair cascades in a charge-limited flow from the stellar surface.
   }
   {
    Our model quantitatively and qualitatively explains the observed pulsar radio powers and spectra, the pulse profiles, polarization curves, their temporal variability, the strong Stokes-$L$ and weak Stokes-$V$ polarization components, the decline in the linear polarization with frequency, and the nonexistence of a radius-to-frequency relation.
   The observable properties of radio emission from the polar cap can vary and include single- or double-peaked profiles.
   Most of the Stokes~$V$ curves from our simulations appear to be antisymmetric, but symmetric curves are also present at some viewing angles.
    Although the polarization-angle (PA) swing of the radiation from the polar cap fits the rotating vector model (RVM) for most viewing angles, the angles obtained from the RVM do not correspond to 
    the dipole geometry of the magnetic field. Instead, the PA is directly related to the plasma flows in the polar cap. 
    Furthermore, we found that the radiation is associated with escaping plasma bunches and can propagate freely along
    channels of low plasma density, in addition to being reflected at the channel boundaries.
   }
   {
    Our simulations demonstrate that pair discharges close to the surface of the polar cap cause the radio emission of pulsars and determine the majority of their typically observed properties. 
The merits of RVM for estimations of the magnetic field geometry from observations need to be reevaluated.
   }


   \maketitle

\section{Introduction} \label{sec:intro}
Most of the coherent radio emission of rotation-powered pulsars is driven by pair cascades and originates in or close to polar caps \citep{Ruderman1975,Cheng1977b}.
The initial characteristics of the coherent radio waves that escape from the polar cap have been the subject of research for many years \citep{Melrose2017a,Melrose2020a,Philippov2022}, but this has not yielded a final and satisfactory model so far.
The initial properties of the emission are, however, essential for the calculation of radiative transfer in pulsar radio-emission models \citep{Petrova2000,Beskin2012,Hakobyan2017,Cao2024a} and any interpretation of the observed polarization \citep{Stinebring1984,Bilous2016,Oswald2023,Cao2024b}.

The pair-discharge cascades in polar caps play an essential role in many plasma processes of pulsars.
They cause the emission of electromagnetic waves over a wide spectral range \citep{Hobs2004,Daugherty1996,Petri2019,Giraud2021}, instabilities resulting in coherent radio emission \citep{Asseo1998,Melikidze2000,Gil2004}, particle acceleration \citep{Daugherty1982}, relativistic plasma outflows \citep{Petri2022}, they fill the neutron star magnetosphere with pair plasma \citep{Tomczak2023}, and they cause neutron star heating \citep{Zhang2000,Gonzales2010,Kopp2023}.

Various coherent radio emission mechanisms in or near pulsar polar caps have been proposed since the discovery of the first pulsar (for a summary, see \citet{Melrose1999,Eilek2016,Melrose2020a} and references therein).
The mechanisms include coherent curvature radiation of charge solitons produced by relativistic streaming instabilities \citep{Melikidze1980,Melikidze2000,Mitra2017,Manthei2021,Benacek2021a,Benacek2021b,Benacek2024,Mitra2024}, relativistic plasma emission by wave-wave interactions \citep{Weatherall1997}, emission by particles undergoing linear acceleration along the magnetic field \citep{Melrose2009a,Melrose2009b,Benacek2023} or in an arbitrary direction to the magnetic field in free-electron maser emission \citep{Fung2004}, anomalous Doppler emission \citep{Lyutikov1999a}, and electron cyclotron or synchrotron maser emission \citep{Kazbegi1991,Labaj2024}.
To date, no decisive observational evidence favoring one of these mechanisms has emerged.

The pulsar magnetosphere is likely to be nonstationary and inhomogeneous. Early attempts to model macroscopic magnetospheric filamentary instabilities were made by \citet{Urpin2014} in order to explain the temporal fluctuations of pulsar radio emission. \cite{Luo2008} showed that temporal oscillations of the polar cap can generate large-amplitude superluminal waves. This mechanism of direct  radio emission generated by the nonstationary plasma during pair discharges has also been simulated by \citet{Philippov2020}.
These authors suggested that fluctuating charge inhomogeneities in the intermittently created pair plasma produce charged bunches and a strong  coherent electric field component classified as superluminal O-mode electromagnetic waves, which then transform into escaping electromagnetic waves when they encounter a drop in plasma density \citep{Philippov2022}.
The wave generation was confirmed by their particle-in-cell (PIC) kinetic simulations \citep{Philippov2020}, which showed broadband spectra of the radiation in a frequency range commensurate with the observed pulsar spectra. The inclusion of nonstationary pair cascades in PIC simulations results in a more physically accurate description of the magnetospheric properties that can also avoid the problems associated with plasma bunching and wave growth that are known to have plagued earlier models \citep{melrose_relativistic_1999}.    

\citet{Chernoglazov2024} studied the cascades in 3D, finding the 3D polar cap structure and coherence scales of the cascades to be on the order of the electric gap height.
The kinetic effects in the polar cap plasma of aligned pulsars were studied by \citet{Cruz2021b} by considering pair cascades along diverging magnetic field lines. 
The authors detected two peaks in the direction of the outward Poynting flux, which were interpreted as being potentially associated with the core and conal components of pulsar radio emissions \citep{Rankin1990,Rankin1993}.
The excitation of electromagnetic waves in pair discharges was also confirmed by global 2D PIC simulations of the aligned rotator's magnetosphere by \citet{Bransgrove2023}, who
demonstrated that electromagnetic waves can also be generated in other magnetospheric regions.
\citet{Tolman2022} showed that after initially strong exponential damping, the damping of these waves becomes linear in time.
The waves can still carry enough energy when they decouple from the plasma in places where the plasma density is low, allowing conversion into vacuum electromagnetic waves that can then be observed as pulsar radio emission.

Using 2D simulations, \citet{Benacek2024b} found that the radio waves generated in pair cascades can be transported away from their origin in the dense polar cap in Poynting flux channels that form in the typical profiles of magnetospheric currents across the polar cap.  
The channels follow the magnetic field lines where the plasma density is low, and they do this as a result of locally small parallel electric fields and currents that do not support the generation of pair cascades.
The plasma frequency in the channels is well below the frequencies of the generated electromagnetic waves, which can then propagate without significant absorption along the field lines.
\citet{Benacek2024b} also showed that the Poynting flux in dense plasma bunches decreases quickly with increasing distance from the star, whereas the flux in the Poynting flux channels shows no significant decrease.
The spectrum of electromagnetic waves in the channels is similar to that of a typical radio pulsar.
The radio waves are expected to leave the magnetosphere, but more detailed modeling of the radiative transfer is required to assess any further modification before the radiation escapes from the magnetosphere \citep{vonHoensbroech1998}.

The detailed properties of the Poynting flux that escapes from the polar cap are essential for specific model predictions about observable radio-emission characteristics, but they remain partly obscured in 2D simulations.
By definition, the polarization vector lies in the plane of the simulation domain for 2D simulations; hence, we cannot obtain any information about the polarization properties from these simulations.
To the best of our knowledge, however, polarization studies for pulsars employing 3D kinetic simulations have not been made to date.

We therefore focused on the polarization properties of the electromagnetic radiation that escapes from the pulsar polar cap as the result of nonstationary pair production in an inhomogeneous polar cap plasma.   
The polarization properties were obtained from scaled 3D fully kinetic PIC simulations that describe the pair discharges and the associated wave--particle interactions.
Pair creation is driven by magnetospheric currents whose profile across the polar cap was obtained from well-known global general-relativistic force-free simulations of the magnetosphere.

This paper is structured as follows.
In Sect.~\ref{sec:methods} we discuss the plasma and numerical setup of the pair cascades in the pulsar polar cap.
Section~\ref{sec:results} describes the plasma properties and shows the polarization properties of the escaping waves when the quasi-periodic nature of the cascades has established itself.
We discuss the obtained properties in Sect.~\ref{sec:discuss} and describe the main polarization properties in Sect.~\ref{sec:conclude}.

\section{Methods} \label{sec:methods}
We studied the polar cap region of pulsars, which has an electric gap zone height comparable with the polar cap radius,
\begin{equation}
    l_\mathrm{gap} \approx r_\mathrm{pc}.
\end{equation}
We represented the nonstationary polar cap by a 3D model in a fully kinetic PIC simulation, augmented by pair cascades, magnetospheric electric currents, an external magnetic field dipole, and particle injections.
Our simulation setup was similar to that developed by \citet{Benacek2024b}, but included a few improvements toward more realistic magnetospheric conditions.
We summarize the main simulation parameters in Table~\ref{tab1}.
The grid cells cannot resolve the realistic kinetic scales of the polar cap, and we therefore downscaled the plasma density and associated quantities to resolve the plasma skin depth and plasma frequency.
The simulated geometry of the polar cap contained open and closed magnetic field lines. The simulation covered several electric gap-zone heights above the star. The electric gap-zone height is defined as the typical acceleration distance of primary particles.
The properties of the escaping radiation were determined in a virtual plane close to the top of the simulation domain, opposite to the stellar surface.

\subsection{Polar cap model}

\begin{figure}[t]
    \includegraphics[width=0.5\textwidth]{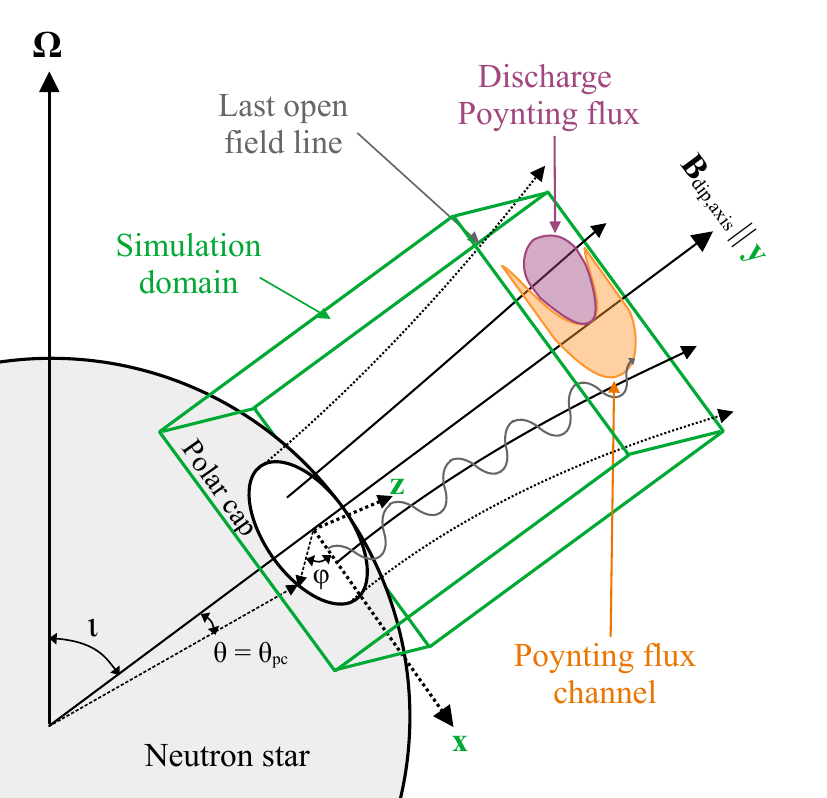}
    \caption{Scheme of the polar cap in the our model (not to scale).
        \label{fig1}
    }
\end{figure}

\begin{table}[htp]
        \centering
    \caption{Parameters for the star and the simulation.}
        \begin{tabular}{lc}
                \hline \hline
        Parameter & Values \\
        \hline
        Dipole inclination & 60$^\circ$ \\
        Dipole magnetic field & 10$^{12}$\,G \\
        Star rotation period $P$ & 0.25\,s \\
        Star mass $M_\star$ & 1.5\,M$_\mathrm{s}$ \\
        Star radius $R_\star$ & 10\,km \\
        \hline
        Domain size $L_x$ & 700$\Delta_\mathrm{x}$ \\
        Domain size $L_y$ & 1500$\Delta_\mathrm{x}$ \\
        Domain size $L_z$ & 700$\Delta_\mathrm{x}$ \\
        Grid cell size $\Delta_\mathrm{x}$ & 1.67\,m \\
        Initial particle density on axis $n_0(\boldsymbol{x} = \boldsymbol{0})$ & $1.4 \cdot 10^{11}\ \mathrm{cm}^{-3}$ \\ 
        Initial plasma skin depth $d_\mathrm{e}'(n_0'(\boldsymbol{x}=\boldsymbol{0})) / \Delta_\mathrm{x}$ & 9.9 \\
        Simulation time step $\Delta t$ & 2.17\,ns \\
        Total simulation time steps $T / \Delta t$ & 20\,000 \\
        Polar cap angle $\theta_\mathrm{pc}$, Eq.~\ref{eq:cap_angle} & 1.97$^\circ$ \\
        Polar cap transition angle $\Delta\theta / \theta_\mathrm{pc}$ & 0.1 \\
        Current profile $j_\parallel / j_\mathrm{GJ,axis}$ & Eq.~\ref{eq:current_profile}, Fig.~\ref{fig2} \\
        Particle decay threshold $\gamma_\mathrm{thr}$ & $1.67\cdot 10^{9}$ \\ 
        Secondary particle $\gamma_\mathrm{sec}$ & $7.3\cdot 10^{4}$\\ 
        \hline
        Initial density $n_0'(\boldsymbol{x}, \iota = 60^\circ)$ & 1\,PPC \\
        Initial and injection particle velocity $u_\mathrm{0}' / c$ & 0.1 \\
        Initial density closed lines $n_\mathrm{closed}'$ & 50\,$n_0'$ \\
        Absorbing boundary condition width $l_\mathrm{abc}$ & 24$\Delta_\mathrm{x}$ \\
                \hline \hline
        \end{tabular}
    \tablefoot{The simulation-scaled quantities are denoted by primes.}
        \label{tab1}
\end{table}

We assumed a spherical neutron star with a radius $R = 10$\,km and a mass $M_\star = 1.5\,M_\mathrm{s}$, where $M_\mathrm{s}$ is the solar mass, rotating with a period $P = 0.25$\,s.
The star has a dipole magnetic field inclined at $\iota = 60^{\circ}$ to the star's rotation axis, $\boldsymbol{\Omega}$, with on-axis strength at the stellar surface ${B}_\mathrm{dip,axis} = {B}_\mathrm{dip}(\iota, R_\star) = 10^{12}$\,G.

The magnetic field is assumed to be composed of two components,
\begin{equation}
    \boldsymbol{B}(\boldsymbol{x},t,\iota) = \boldsymbol{B}_\mathrm{dip}(\boldsymbol{x},\iota) + \delta \boldsymbol{B}(\boldsymbol{x},t),
\end{equation}
where $\delta \boldsymbol{B}$ is the local space- and time-varying field, and $\boldsymbol{B}_\mathrm{dip}$ is the external dipole field, which does not evolve in time.

Following \citet{Gralla2017}, we use the Euler potential for the last open magnetic field line,
\begin{equation}
    \alpha_\mathrm{pc}(\iota) = \sqrt{\frac{3}{2}} \mu \Omega \left( 1 + \frac{1}{5} \sin^2 \iota \right),
\end{equation}
where $\Omega = 2\pi / P$ is the angular frequency and $\mu$ is the dipole magnetic moment of the star, to specify the polar cap radius. 
The polar cap is described in spherical coordinates $(\theta,\varphi, r)$ centered on the star, with the polar angle $\theta$ being measured from the magnetic dipole axis and the azimuthal angle $\varphi$ being counted from the meridian that passes through the dipole axis as shown in Fig.~\ref{fig1}.
Then, the last open magnetic field line corresponds to the polar cap angle, 
\begin{equation} \label{eq:cap_angle}
\theta_\mathrm{pc}(\iota) \equiv \arcsin  \sqrt{\frac{\alpha_\mathrm{pc}R_\star}{\mu}} \approx 1.97^{\circ}.
\end{equation}
There is also a transition angle $\theta \in (\theta_\mathrm{pc},\theta_\mathrm{pc} + \Delta \theta)$ between the closed and open field lines, where $\Delta \theta = 0.1 \, \theta_\mathrm{pc}$.

\subsection{Magnetospheric currents}
The global magnetospheric current densities, $j_\mathrm{mag}(\boldsymbol{x})$, $\boldsymbol{x} = (x,y,z)$, modify the Maxwell field solver as described in \citet[Appendix~A]{Timokhin2010},
\begin{equation} \label{eq:electric_field}
    \frac{\partial \boldsymbol{E}(\boldsymbol{x},t)}{\partial t} = -4\pi \left(\boldsymbol{j}(\boldsymbol{x},t) - \boldsymbol{j}_\mathrm{mag}(\boldsymbol{x}) \right) + c (\nabla \times \delta\boldsymbol{B}(\boldsymbol{x},t)).
\end{equation}
The term $\nabla \times \delta\boldsymbol{B}$ represents the time-evolving part of the magnetic field.
The current density, $j(\boldsymbol{x},t)$, is the result of plasma motion, and the term $j_\mathrm{mag}(\boldsymbol{x})$ is the current density necessary to support the twist of the magnetic field lines in the force-free magnetosphere.
We assume that the magnetospheric current, $\boldsymbol{j}_\mathrm{mag}$, is always parallel to the dipole magnetic field.
The magnetospheric currents at open field lines are expressed
as the analytical fit across the polar cap found by \citet{Gralla2017} and \citet{Lockhart2019} in general relativistic force-free simulations,
\begin{equation} \label{eq:current_profile2}
    \boldsymbol{j}_\mathrm{mag} = \frac{\Lambda}{\sqrt{\Upsilon}} \boldsymbol{B},
\end{equation}
\begin{multline} \label{eq:current_profile}
    \Lambda = \mp 2 \Omega \left\{ J_0\left(2 \arcsin \sqrt{\frac{\alpha}{\alpha_\mathrm{pc}}}\right) \cos \iota  \right. \\
    \mp \left. J_1\left(2 \arcsin \sqrt{\frac{\alpha}{\alpha_\mathrm{pc}}}\right) \cos \beta \sin \iota \right\}, \qquad
    \alpha < \alpha_0,
\end{multline}
where
\begin{equation}
    \alpha = \frac{\mu}{r} \sin^2 \theta', \qquad \beta = \varphi',
\end{equation}
and
\begin{equation}
    \Upsilon \approx 1 - \frac{2 G M_\star}{c^2} \frac{1}{r},
\end{equation}
is the reddening factor.
$J_0$ and $J_1$ are Bessel functions of the first kind, the Euler potentials $(\alpha, \beta)$ are expressed in spherical coordinates $(\theta', \varphi', r)$, and
$G$ is the gravitational constant.
The $\mp$ signs in Eq.~\ref{eq:current_profile} correspond to the north and south magnetic poles, respectively.
Our simulations are for the north pole.
We neglect the effects of retardation and aberration in the magnetic field structure and follow 
\citet{Gralla2017}, who estimated the current profile using symmetry arguments, which lead to a circular shape of the current distribution and the polar cap boundary.

\begin{figure}[t]
    \includegraphics[width=0.45\textwidth]{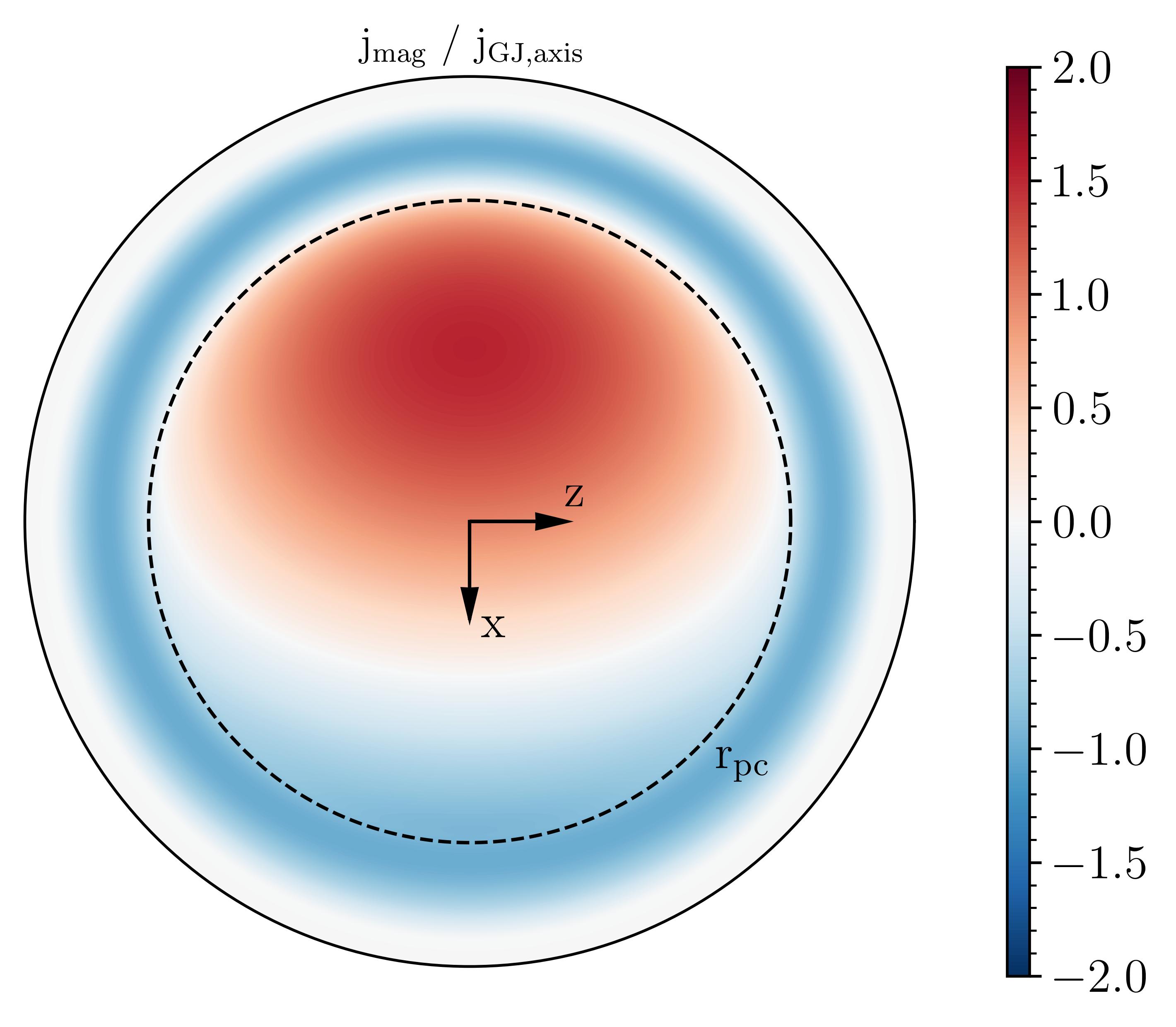}
    \caption{Magnetospheric current profile across the polar cap at the stellar surface.
        \label{fig2}
    }
\end{figure}

We also implement an appropriate return current on the closed field lines to obtain a zero net magnetospheric current over the polar cap.
The return current layer has a half sine profile, and its amplitude is fixed to $j\sqrt{\Upsilon}/j_\mathrm{GJ,axis} = 1$.
The width of the return current is then adjusted so that the resulting net current is zero.
If the magnetospheric current, Eq.~\ref{eq:current_profile2}, is nonzero at the last open field line, the sinusoidal profile is cut to match it at the boundary, resulting in a smooth function of the current.
The current profile across the polar cap is shown in Fig.~\ref{fig2}.

The plasma density is initialized to have the Goldreich--Julian charge density \citep{Goldreich1969}, 
\begin{equation}
    \rho_\mathrm{GJ}(\boldsymbol{x},\iota) \equiv - \frac{\boldsymbol{\Omega} \cdot \boldsymbol{B_\mathrm{dip}}(\boldsymbol{x},\iota)}{2 \pi c},
\end{equation}
assuming to be flowing relativistically and thus yielding the Goldreich--Julian current density for an arbitrary inclination angle and a position vector as
\begin{equation}
    j_\mathrm{GJ}(\boldsymbol{x},\iota) \equiv - c \rho_\mathrm{GJ}(\boldsymbol{x},\iota).
\end{equation}
Ignoring general relativistic corrections and for simplicity, we normalize the current and number densities on parameters found in the magnetic axis of an aligned pulsar as
\begin{eqnarray}
    j_\mathrm{GJ,axis} & \equiv & j_\mathrm{GJ}(\boldsymbol{x}=\boldsymbol{0},\iota=0), \\
    n_\mathrm{GJ,axis} & \equiv &\rho_\mathrm{GJ}(\boldsymbol{x}=\boldsymbol{0},\iota=0) / e \approx 2.8\times 10^{11} \mathrm{cm}^{-3}.
\end{eqnarray}

\subsection{Particle-in-cell simulations}
We investigate the polar cap in a frame co-rotating with the neutron star.
The plasma is described with a fully kinetic, relativistic  
3D3V version of the electromagnetic PIC code ACRONYM \citep{Kilian2012}.

We use the fourth-order finite-difference time-domain (FDTD) method for computation on the Yee lattice to efficiently describe wave dispersion in relativistic plasmas.
For the current deposition, we apply the \citet{Esikperov2001} current-conserving deposition scheme with a fourth-order shape function of macro particles \citep{Lu2020}.
The synchrotron loss time in the polar cap region is much shorter than all other timescales in our simulation, leading to ignorable values of the momentum component perpendicular to the magnetic field, $p_\perp$. Hence we use the \citet{Vay2011} particle pusher in gyro-motion approximation  as described by \citep{Philippov2020}, for solving with a leapfrog algorithm the guiding centre equations for the particle momentum parallel to the magnetic field $\frac{\mathrm{d}\boldsymbol{p_{||}}}{\mathrm{d}t}=e\boldsymbol{E}_{||}$
and its position $\frac{\mathrm{d\boldsymbol{x}}}{dt}=v_{||}\frac{\boldsymbol{B}}{B}+c\boldsymbol{E}_{\perp}\times\frac{\boldsymbol{B}}{B^{2}}$.

The simulation has a uniform rectangular grid of size $(L_\mathrm{x} \times L_\mathrm{y} \times L_\mathrm{z}) = (700\Delta_\mathrm{x} \times 1500 \Delta_\mathrm{x} \times 700 \Delta_\mathrm{x})$, where $\Delta_\mathrm{x} \approx 1.67$\,m is the grid cell size, giving the real size of the domain $\approx$1170\,m $\times$ 2500\,m $\times$ 1170\,m.
The grid cell size resolves distances smaller than the electric gap size of $l_\mathrm{gap} \simeq 100$\,m, and the scaling $L_\mathrm{y} \gg l_\mathrm{gap} \gg \Delta_\mathrm{x}$ holds.
The time step is chosen as $\Delta t = 0.45 \Delta_\mathrm{x}/c \approx 2.17$\,ns.
We conduct the simulation for 20\,000 time steps, corresponding to 43.4\,$\mu$s.

The orientation of the coordinate axes is illustrated in Fig.~\ref{fig1}.
The $y$-axis of the simulation domain is along the dipole axis, and the $x-y$ plane is in the $\boldsymbol{\Omega} - \boldsymbol{B}_\mathrm{dip,axis}$ plane.
The center of the coordinate system ($x=0$, $y=0$, $z=0$) is on the dipole axis, where the stellar surface ($r = R_\star$) intersects at $\approx 55 \Delta_\mathrm{x}$ above the bottom of the simulation grid.

\subsection{Boundary and initial conditions}
We assume that the initial electric field is zero, and the initial magnetic field $B_\mathrm{ext}$ is given only by the dipole magnetic field.
The open magnetic field lines are initially filled with $n_0 = 1$ macro-particles per cell (PPC), providing the Goldreich-Julian charge density.
Thus, the plasma in open magnetic field is not dense enough to fully sustain the magnetospheric currents. The closed field lines are filled by a new pair when their number density drops below $n_\mathrm{closed} = 5$\,PPC, half electrons and half positrons.
Each macro-particle represents 50$n_\mathrm{GJ,axis}$.
Between these regions we implemented a linear transition in density over the angular width $\Delta \theta$.

All particles have an initial Maxwell-Boltzmann distribution function with a thermal velocity $u_\mathrm{0} = 0.1\,c$.
The simulation has open boundary conditions for particles.
We inject particles into the simulation at every time step.
At open field lines, we inject particles at the stellar surface and at the ``top'' boundary far from the star ($y = L_\mathrm{y}$), if the number density drops below a threshold of $n_0$.
The injection happens in a layer with a thickness of one grid cell.
At closed magnetic field lines the threshold for injection is $n_\mathrm{closed}$. If the macro-particle number falls below the threshold, an
electron-position pair is injected at the simulation boundaries and an electron-proton is injected at the stellar surface, which maintains charge neutrality.

We applied the algorithm called complex shifted coefficient --- convolutionary perfectly matched layer (CFS---CMPL) \citep[Chapter~7]{Roden2000,Taflove2005} in the ghost cells surrounding the simulation domain as absorbing boundary conditions for electromagnetic waves.
We inject 10\,PPC below the stellar surface, each with ten times the macro-factor of particles at open field lines.
These particles rearrange themself to screen the electric field.
Particles were removed when they reached the simulation boundary or when they crossed the stellar surface to maintain the charge-limited flow condition.
Additional particles were injected when the density drops below 10\,PPC.

We tested a simulation with the same grid resolution and an electron--positron surface injection.
That leads to a filling with positrons of the field lines with $j_\mathrm{mag}/j_\mathrm{GJ} < 0$ and to changes of the density structure.
However, Poynting-flux channels are still formed in low-density channels and remain sparse outside the channels.
The polarization properties of the radiation
also remain mostly unchanged.

\subsection{Scaling of the plasma quantities}
\label{ssect:scaling}

\begin{table}[tp]
        \centering
\caption{Scaling of the physical quantities in the simulations.}
        \begin{tabular}{ll}
                \hline \hline
        Parameter Name & Simulation Scaling \\
        \hline
        Time & $t' = t$ \\
        Position vector & $\boldsymbol{x}' = \boldsymbol{x}$ \\
        Light speed & $c' = c$ \\
        \hline
        Plasma density & $n = \zeta n'$ \\
        Plasma frequency & $\omega_\mathrm{p} =  \zeta^{\frac{1}{2}} \omega_\mathrm{p}'$ \\
        Plasma skin depth & $d_\mathrm{e} = d_\mathrm{e} / \zeta^{\frac{1}{2}}$ \\
        Electric current density & $j = \zeta j'$ \\
        Electric field intensity & $E = \zeta E'$ \\
        Magnetic field intensity & $B = \zeta B'$ \\
        Poynting flux & $S = \zeta^2 S'$ \\
        Threshold Lorentz factor & $\gamma_\mathrm{thr} = \zeta \gamma_\mathrm{thr}'$ \\
        Secondary  particle Lorentz factor & $\gamma_\mathrm{sec} = \sqrt{\zeta} \gamma_\mathrm{sec}'$ \\
                \hline \hline
        \end{tabular}
    \tablefoot{The simulation-scaled quantities are denoted by primes. The scaling factor is $\zeta$.}
        \label{tab2}
\end{table}

As the simulation grid size does not resolve the electron skin depth of the real plasma at the polar cap, the plasma parameters associated with the skin depth are scaled by a factor $\zeta$,
\begin{equation} \label{eq:scaling}
    \zeta \equiv \left(\frac{d_\mathrm{e,real}}{d_\mathrm{e,simulation}'}\right)^2 = \frac{j_\mathrm{real}}{j_\mathrm{simulation}'} \approx 1.33\times10^6.
\end{equation}
For more details, see Table~\ref{tab2}.
The scaled quantities used in the simulation are denoted by primes.

The scaled initial plasma density resolves the skin depth by about 20$\Delta_\mathrm{x}$.
When the plasma density reaches a local maximum in a bunch, the skin depth is still resolved by $\gtrsim$2$\Delta_\mathrm{x}$.
The scaling of Eq.~\ref{eq:scaling} also applies to magnetospheric currents and electromagnetic fields because in Eq.~\ref{eq:electric_field} the scaled magnetospheric currents influence the electric and magnetic fields.

\subsection{Pair-cascade simulation}
For the pair cascade, we adopt the algorithm by \citet{Philippov2020,Cruz2021a,Cruz2022} based on a threshold Lorentz factor, $\gamma_\mathrm{thr}$, for pair-production.
In that approach, the quantum electrodynamic process is approximated by the production of a new electron--positron pair whenever the Lorentz factor of the primary particle exceeds $\gamma_\mathrm{thr}'$.
We define a threshold Lorentz factor $\gamma_\mathrm{thr} = \zeta \gamma_\mathrm{thr}' = 1.7\times10^{9}$.
By scaling the threshold $\gamma_\mathrm{thr}'$ as $\sim \zeta^{-1}$ we ensure that the distance in which a charged particle is accelerated to the pair-creation threshold in the real polar cap is the same in the simulated polar cap (for a uniform and constant electric field).
Hence, the ratio between the acceleration distance and the polar cap radius remains the same as for the unscaled polar cap.

The macro-particles suffer losses by a radiative-reaction force for curvature-radiation of the form \citep{Jackson1998,Daugherty1982,Timokhin2010,Tamburini2010}
\begin{equation} \label{eq-RL}
    \left(\frac{\mathrm{d}p}{\mathrm{d}t}\right)_\mathrm{RR} = \frac{2 q^2}{3 m c} \frac{p^4}{\rho^2},
\end{equation}
where $p = \gamma \beta$ is the dimensionless particle momentum, $q$ is the particle change, $m$ is the particle mass, and $\rho$ is the curvature radius.
We choose a constant curvature radius of $\rho = c / \Omega = 10^{7}$\,cm.
Although the magnitude of the curvature radius influences both the emission of gamma-ray photons by curvature radiation and the photon mean free path, in our simplified model it only affects the radiation losses given by Eq.~\ref{eq-RL}.
In reality the curvature radius will also vary across and along the polar cap and may somewhat modify the overall picture by a certain amount. The general morphology is, however, expected to be similar to that in our simplified simulation scenario.

New pairs are removed whenever the density in a given grid cell exceeds 120\,PPC.
This condition decreases the computational imbalance between processors and also ensures that the skin-depth resolution remains sufficiently high in the regions with the highest density.
Our tests show that the volume of regions, where the density would exceed the maximum number density, is negligibly small in comparison to the total domain volume.

\section{Results}  \label{sec:results}

\begin{figure*}[ht!]
    \centering
    \includegraphics[width=0.4\textwidth]{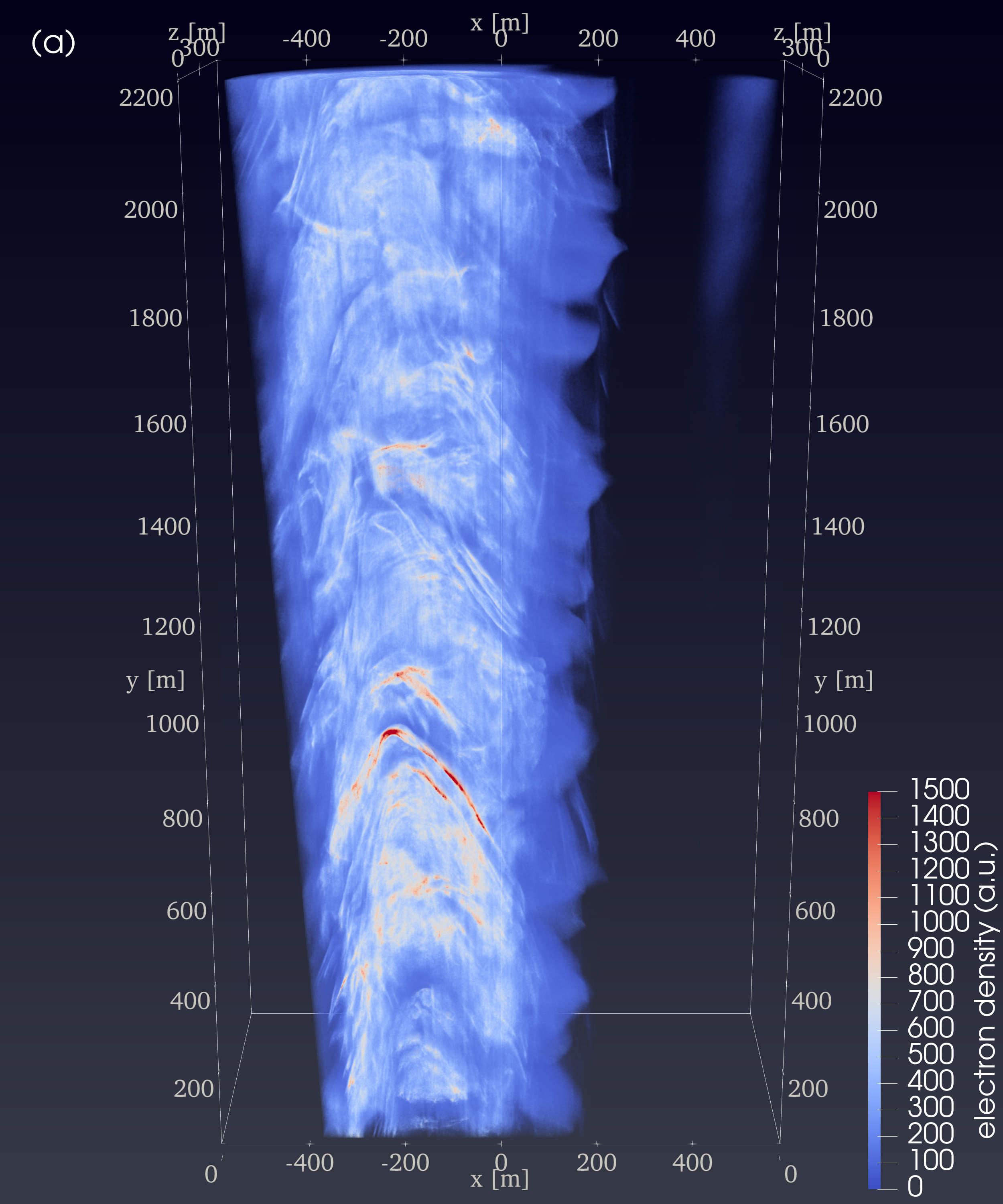}
    \includegraphics[width=0.4\textwidth]{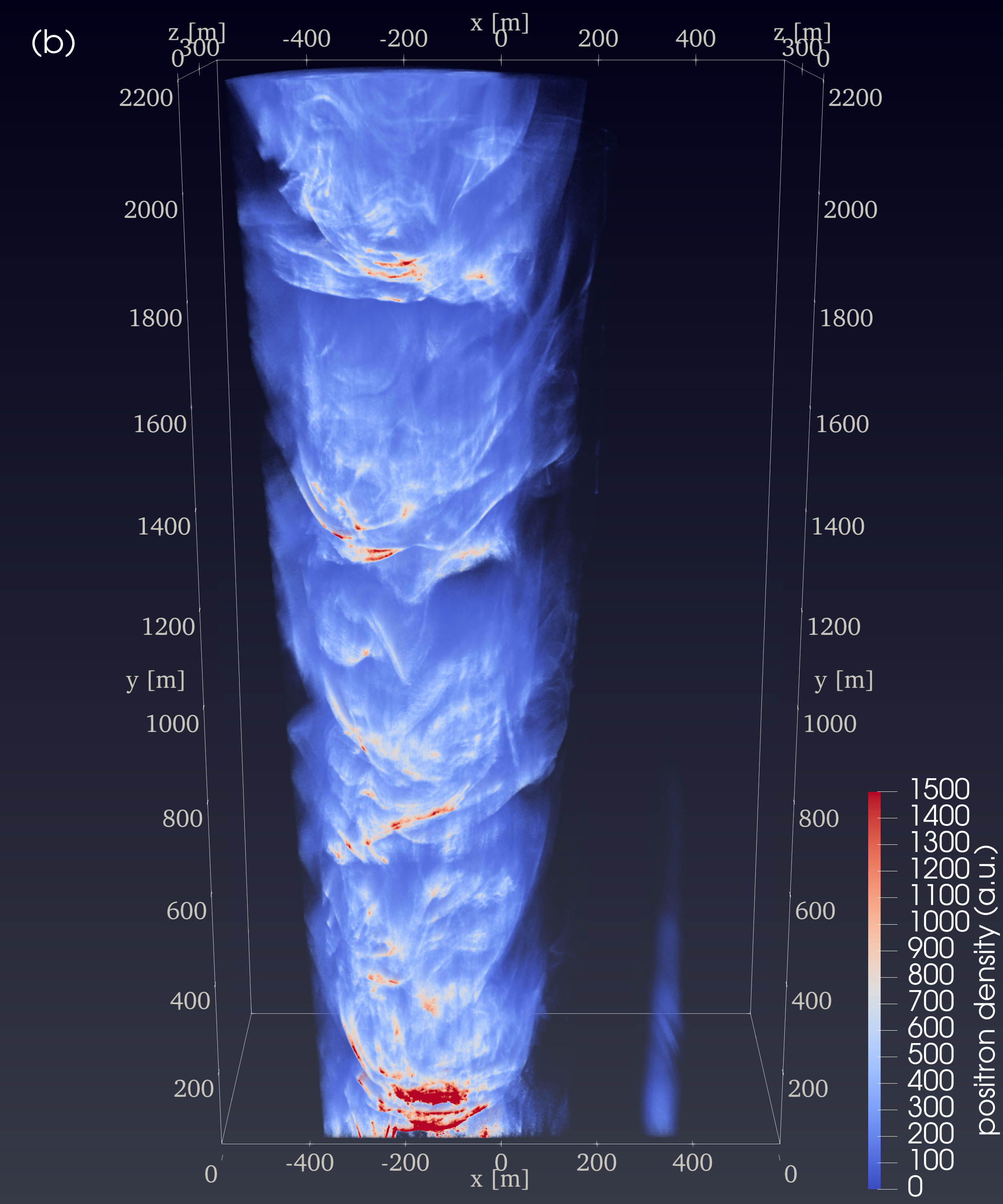} \\
    \includegraphics[width=0.4\textwidth]{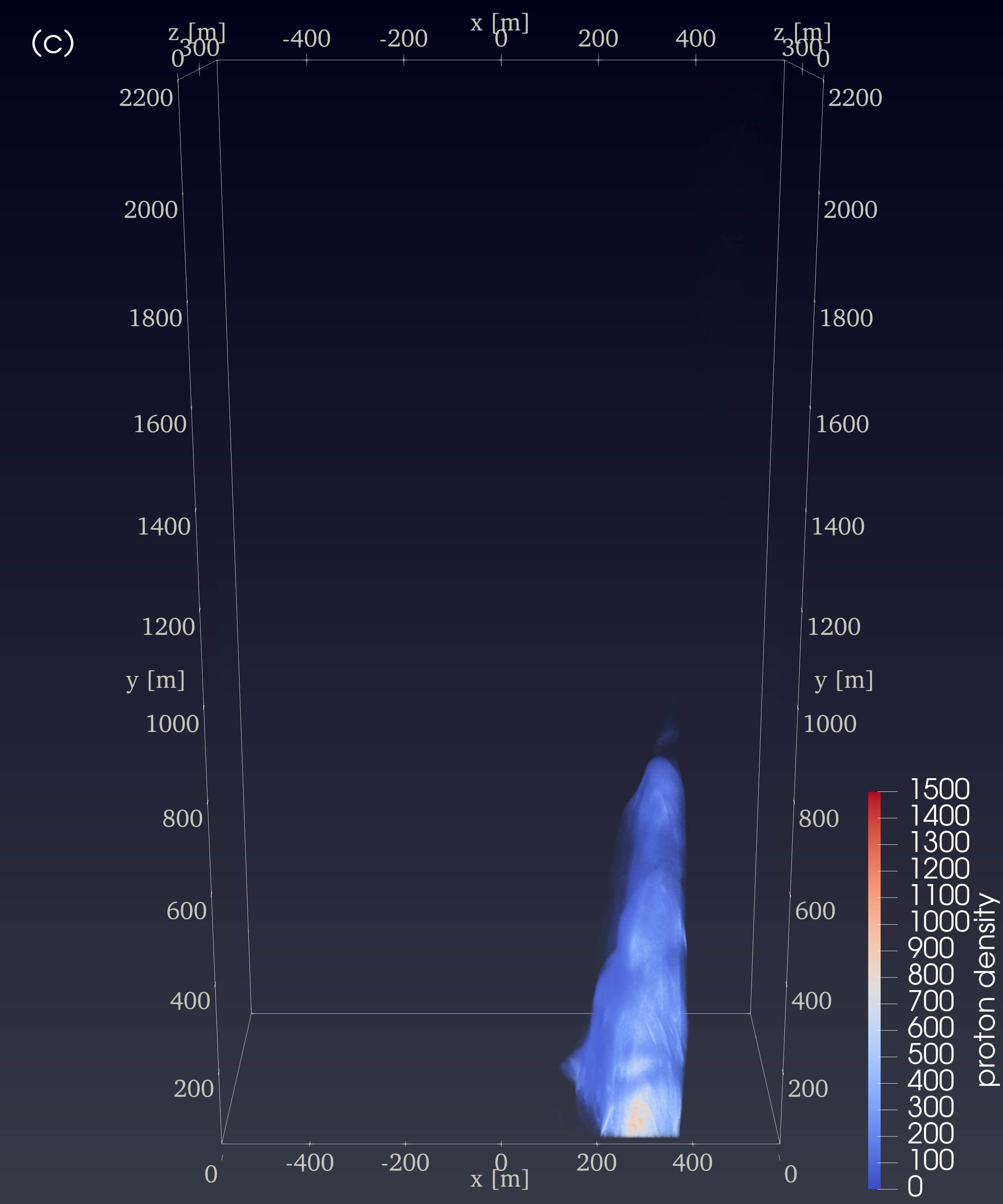}
    \includegraphics[width=0.4\textwidth]{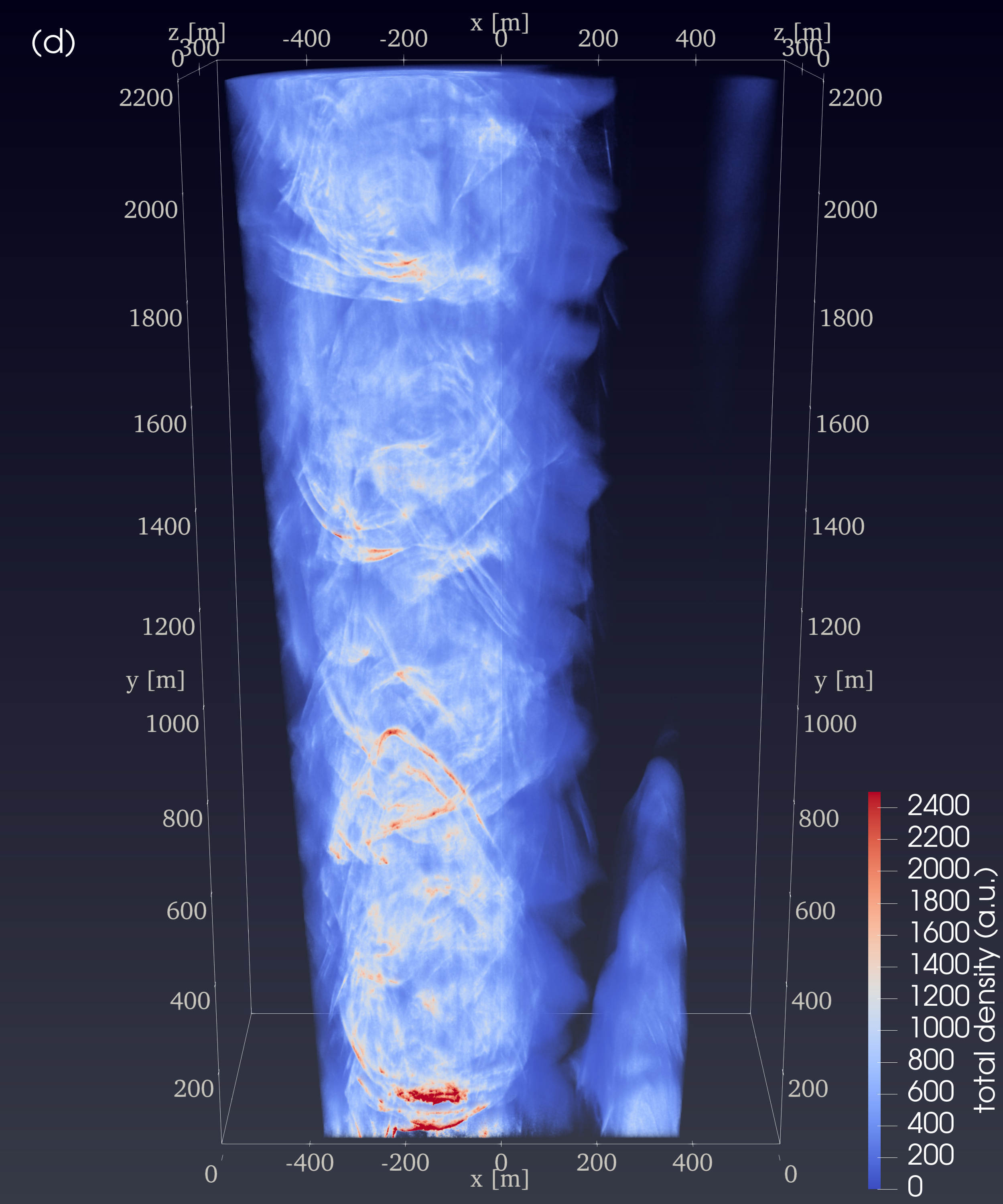}
    \caption{Plasma number density in the polar cap at the simulation end.
        (a) Electron density, (b) positron density, (c) proton density, and (d) total density.
        One half of the domain is selected for $z > 0$.
        The quantities in closed field lines are set to zero.
        }     
        \label{fig3}
\end{figure*}

\begin{figure*}[ht!]
    \centering
    \includegraphics[width=0.4\textwidth]{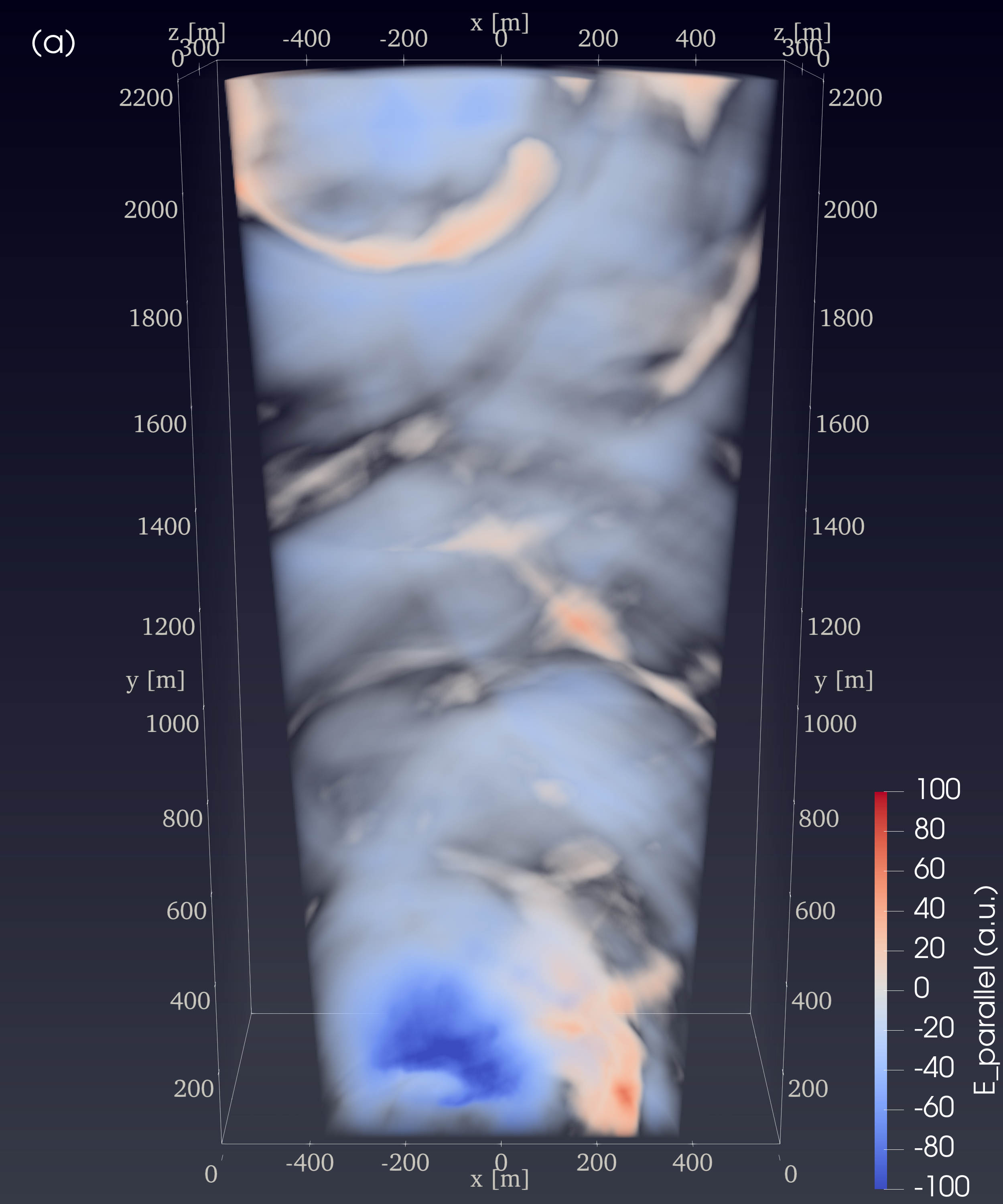}
    \includegraphics[width=0.4\textwidth]{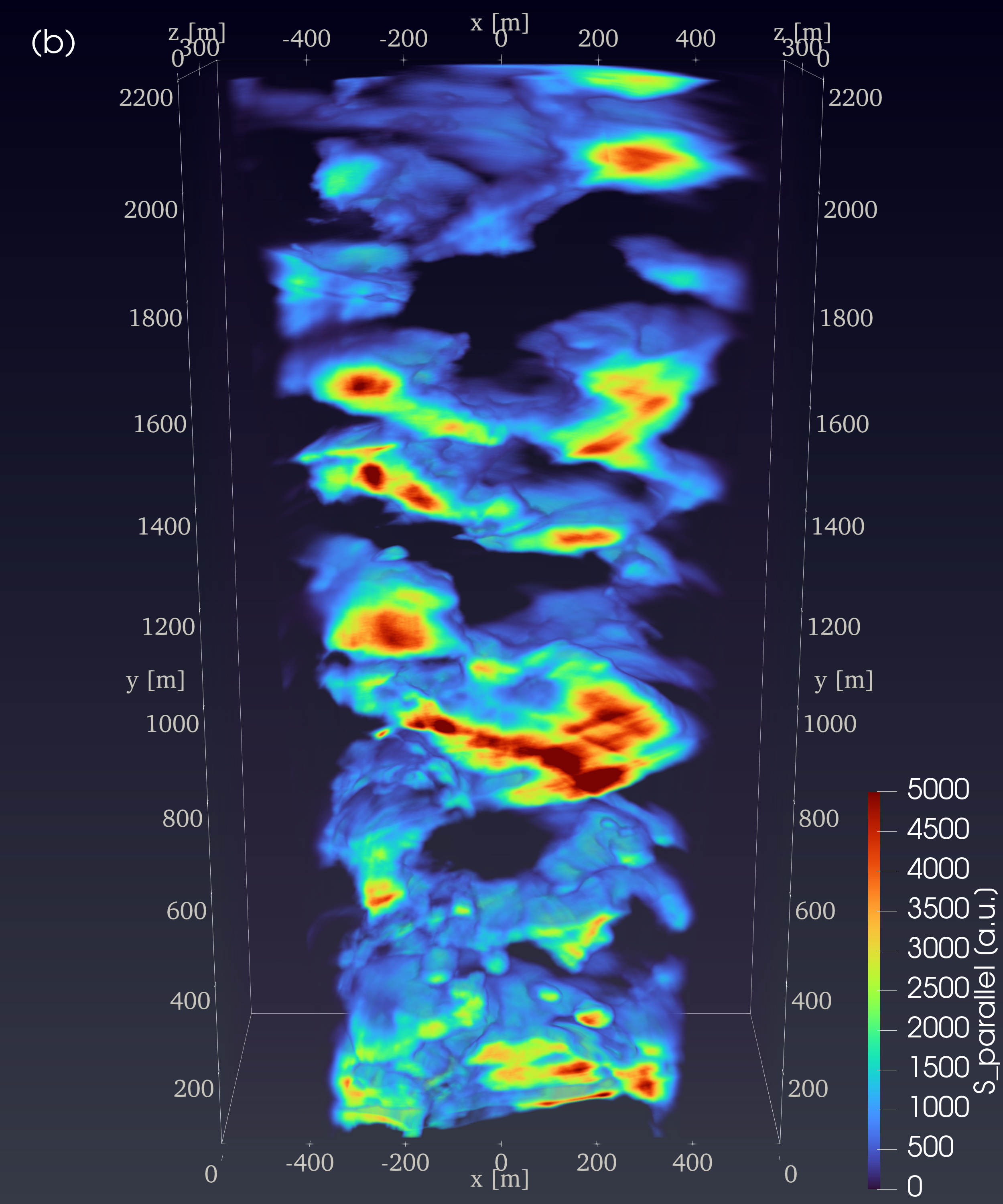}
    \caption{Same as Fig.~\ref{fig3}, but for parallel electric fields (a) and Poynting flux (b), both parallel to the magnetic field.
        }     
        \label{fig4}
\end{figure*}

\begin{figure*}[t!]
    \centering
    \includegraphics[width=0.9\textwidth]{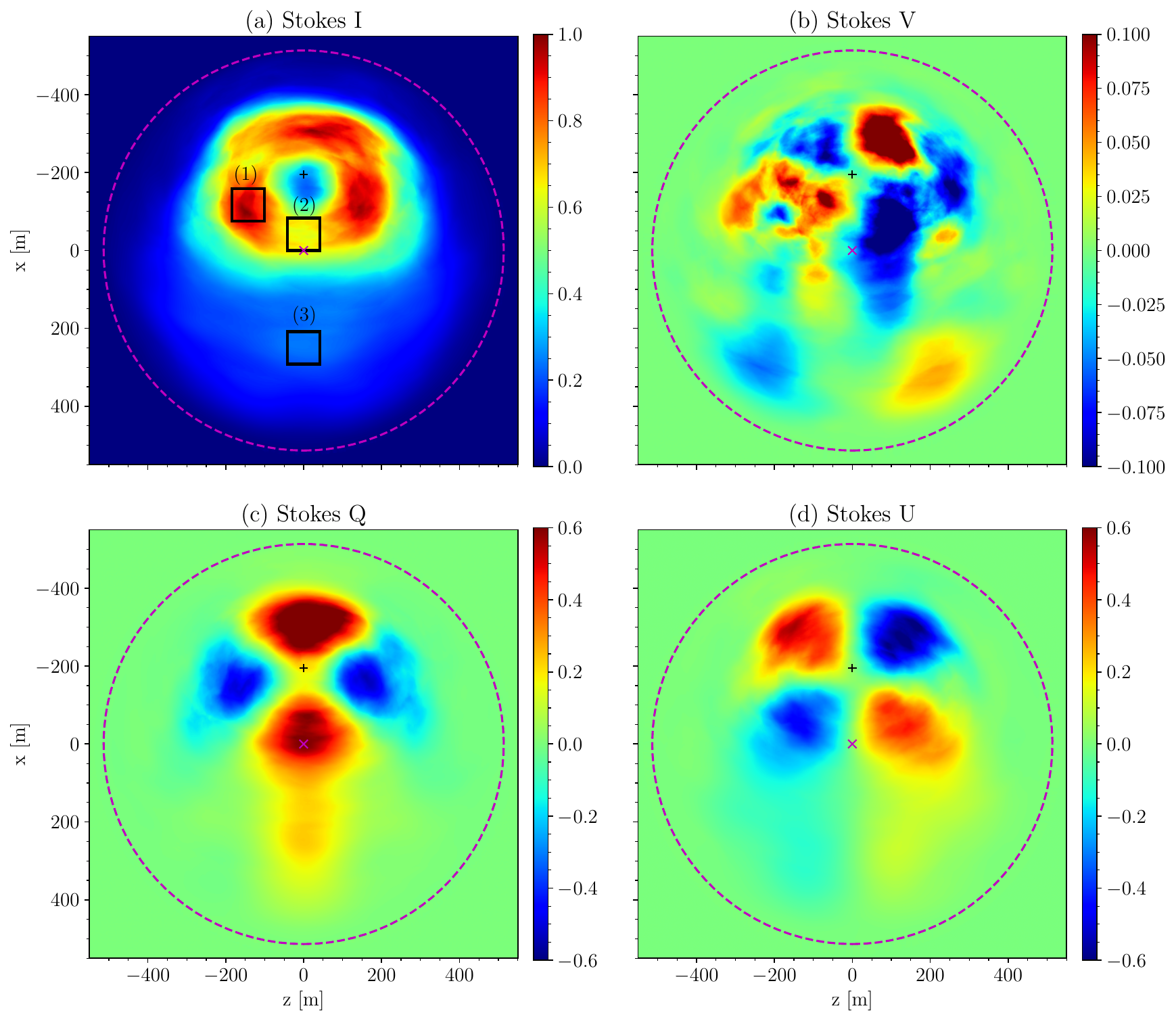}
    \caption{Stokes parameters $I$, $V$, $Q$, and $U$ (a--d)  of the escaping radiation captured in a plane at a height $H \approx 2150$\,m, averaged over the time interval $T \in [34.7,43.4]\,\mu$s, and normalized to the maximum value of the Stokes~$I$ value.
        The dashed magenta line denotes the last open field line, and the magenta plus shows the dipole axis.
        The black plus denotes the maximum of the average plasma density.
        The black squares in (a) denote regions for which the spectra are analyzed below.
        The color scales differ between figures.
        }     
        \label{fig5}
\end{figure*}

\begin{figure}[t!]
    \centering
    \includegraphics[width=0.49\textwidth]{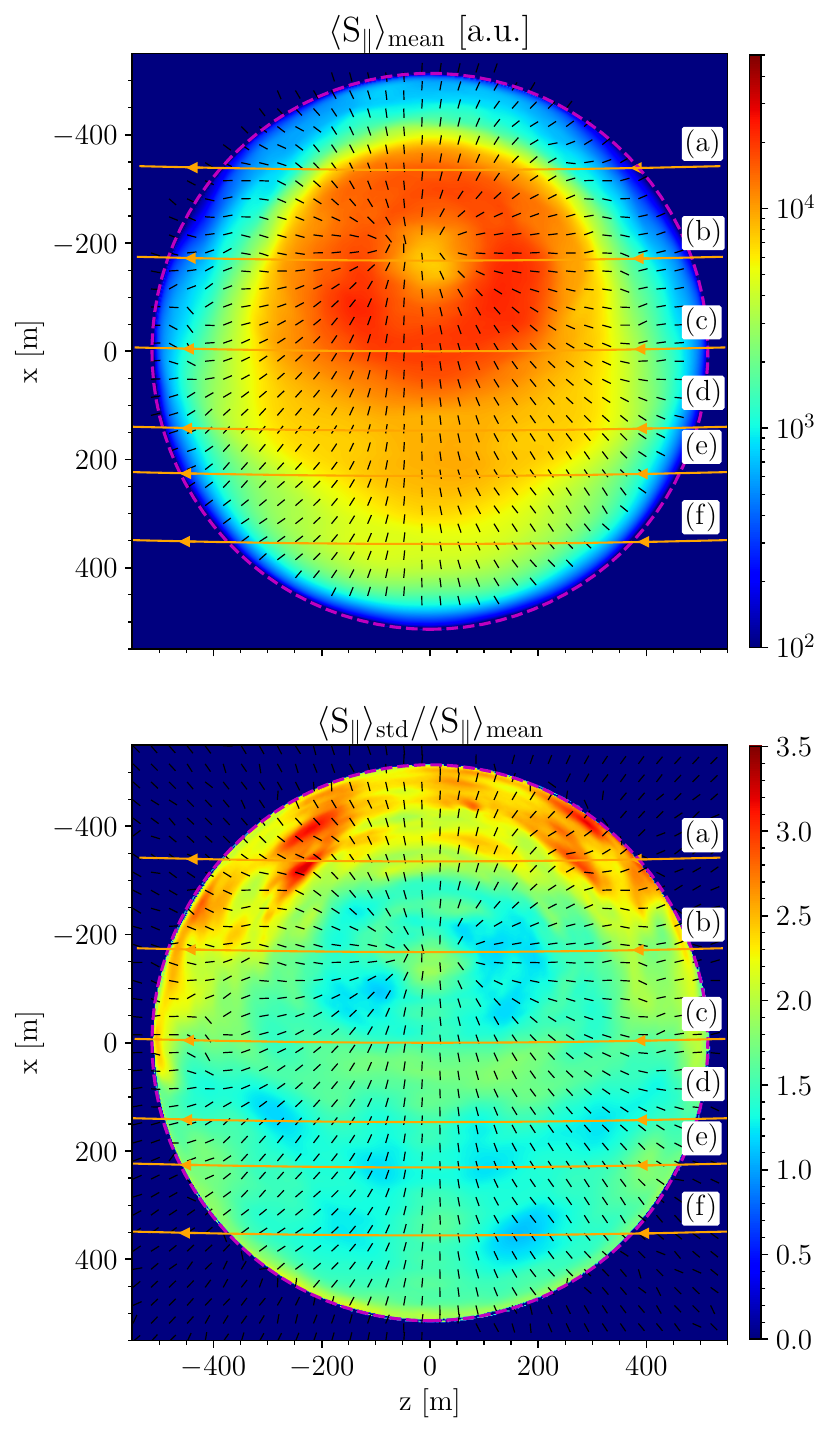}
    \caption{Parallel component of the escaping Poynting flux as captured in the detection plane at a height $H \approx 2150$\,m and averaged over the time interval $T \in [34.7,43.4]\,\mu$s.
        Top panel: Averaged Poynting flux.
        Bottom panel: Normalized standard deviation of the Poynting flux, overlaid with black lines indicating the PA.
        The dashed magenta line denotes the last open field line.
    The orange lines denote the trajectories of constant latitudes used for further investigation.
        }     
        \label{fig6}
\end{figure}

We aim to investigate the electromagnetic polarization of the radio emission from the pair-cascade discharges in the gap that 
become evident when the simulation settles into a quasi-periodic discharge formation close to the stellar surface.
Parallel electric fields start to grow in the first phase of the simulation. They are generated by the magnetospheric currents, according to Eq.~\ref{eq:electric_field}.
The plasma flows out from the polar cap, and pair cascades form at the stellar surface and also close to the upper boundary, filling the polar cap with new pairs.
In the second phase of the simulation the plasma begins to mix along the magnetic field lines and subsequently the repetitive pair cascades begin to form close to the stellar surface.

The quasi-periodic stage is reached after $\approx 3-4L_\mathrm{y}/c$ (12\,000--16\,000 time steps).
Our simulation is set up to quickly and efficiently pass through the initial stages.
In the following sections, we analyze the plasma and field properties after the simulation reached   the quasi-periodic cascade stage, that is, for time steps $\geq$16\,000.

\subsection{Plasma properties of the quasi-periodic discharge}
As the escaping radiation is shaped by the plasma properties, we present an overview of the plasma properties first.
Figure~\ref{fig3} shows the plasma densities of all three species in the polar cap at the end of the simulation, excluding  the regions with closed field lines for greater clarity.
The faint gray box denotes the rendered half-domain for $z>0$.

Most of the electrons and positrons are located in the regions of large and positive magnetospheric currents, $j_\mathrm{mag}/j_\mathrm{GJ} > 0$ in the $x < 0$ part of the domain above the electric gap.
The ions are located only in regions in which the magnetospheric currents have the opposite polarity, $j_\mathrm{mag}/j_\mathrm{GJ} < 0$, which allows ion extraction from the stellar surface.
The pair cascades occur at open field lines where $j_\mathrm{mag}/j_\mathrm{GJ} > 1$ or $j_\mathrm{mag}/j_\mathrm{GJ} < 0$, in agreement with \citet{Timokhin2013}.
At the remaining open field lines, the plasma can sustain the magnetospheric currents, and the plasma density remains low in comparison to the regions in which the pair cascades occur.

Figure~\ref{fig4} shows maps of the electric field and the Poynting flux, both projected onto the magnetic field vector as
\begin{equation}
    E_\mathrm{\parallel} = \frac{\boldsymbol{E} \cdot \boldsymbol{B}}{|\boldsymbol{B}|}, \qquad
    S_\mathrm{\parallel} = \frac{\boldsymbol{S} \cdot \boldsymbol{B}}{|\boldsymbol{B}|}, \qquad
    \boldsymbol{S} = \frac{c}{4\pi}(\boldsymbol{E} \times \boldsymbol{B}).
\end{equation}
The strongest parallel electric fields are located close to the stellar surface, forming an electric gap with a height of $l_\mathrm{gap}$$\approx$400\,m.
Above this gap, the parallel electric field is weaker but still shows structures related to the plasma bunches.

The 
escaping Poynting flux is slightly oblique with respect to the magnetic field,
the radiation being emitted into a narrow angle around the field lines, which can be approximately estimated by $1/\gamma_\mathrm{sec}'$.
Hence, the projection of the flux onto the magnetic field lines determines the strength of the emission.
Most of the perpendicular Poynting flux component
\begin{equation}
    \boldsymbol{S}_\perp = \boldsymbol{S} - \boldsymbol{S}_\parallel
\end{equation}
does not escape from the polar cap and remains localized in the gap region.

The strongest parallel Poynting flux is found close to  the boundaries of the plasma bunches and on  field lines that carry small magnetospheric currents (see Fig.~\ref{fig2}b), similarly to what has been found by \citet{Benacek2024b}.
There is also some negative Poynting flux propagating toward the star, which is not shown in Fig.~\ref{fig3}b.

\subsection{Polarization properties of the electromagnetic waves}
Figures~\ref{fig5} to \ref{fig7} show time- and frequency-averaged profiles of flux and polarization of the electromagnetic radiation that escapes from the polar cap.
The magenta circles in the four panels of Fig.~\ref{fig5} indicate the last closed field lines at the height of the virtual receiving plane, as they cross a plane at a height of $H\approx2160$\,m $\approx 5 l_\mathrm{gap}$ above the stellar surface.
The waves are analyzed at every second time step for time steps 16\,000--20\,000 ($T \in [$34.7, 43.4]\,$\mu$s).
We 
project their wave vector
to the dipole magnetic field for our analysis as these represent the escaping transversal waves and calculate the Stokes vectors in the Fourier space.
Because most of the escaping waves have small angles to the magnetic field, we assume that the amplitude of the escaping waves and their projection to the magnetic field are approximately the same.

\subsubsection{Definition of Stokes parameters}
The frequency-averaged Stokes vector components $(I, Q, U, V)$ on the virtual plane are presented in Fig.~\ref{fig5}.
They are obtained using the FX\footnote{``F'' stands for the Fourier transform and ``X'' for correlation, in this order.} technique \citep[Chapter 6]{Robishaw2018}
\begin{eqnarray}
    I &=& \Bigl\langle E_\mathrm{x}(\omega) E_\mathrm{x}^\ast(\omega)\Bigl\rangle_{\omega} \,+\, \Bigl\langle E_\mathrm{z}(\omega) E_\mathrm{z}^\ast(\omega)\Bigl\rangle_{\omega}, \\
    Q &=& \Bigl\langle E_\mathrm{x}(\omega) E_\mathrm{x}^\ast(\omega)\Bigl\rangle_{\omega} \,-\, \Bigl\langle E_\mathrm{z}(\omega) E_\mathrm{z}^\ast(\omega)\Bigl\rangle_{\omega}, \\
            U &=& \,\,\, 2 \, \mathrm{Re}\, \biggl\{ \Bigl\langle E_\mathrm{x}^\ast(\omega)E_\mathrm{z}(\omega)\Bigl\rangle_{\omega} \biggr\}, \\
        V &=& - 2 \, \mathrm{Im}\, \biggl\{ \Bigl\langle E_\mathrm{x}^\ast(\omega)E_\mathrm{z}(\omega)\Bigl\rangle_{\omega} \biggr\},
\end{eqnarray}
where the components $E_\mathrm{x}(\omega)$ and $E_\mathrm{z}(\omega)$ are the Fourier transformations of the time evolving electric field components $E_\mathrm{x}(t)$ and $E_\mathrm{z}(t)$ for each grid position at the virtual plane projected to the magnetic field.
The asterisk denotes the complex conjugate, $\langle \ldots \rangle_\mathrm{\omega}$ denotes the average over frequency of the Fourier power, which can be understood as the time-average of the real energy density.
The zero frequency component, $\omega=0$, of the Fourier transformation is not included because it does not represent waves.
Our simulation I/O strategy allows us to resolve frequencies in the range $\omega/\omega_\mathrm{p}(n_\mathrm{GJ,axis}) \approx 0.11 - 110$.
The Stokes parameters are computed as the result of frequency averaging of the electric field components with Stokes $U$ and $V$ simply given as the real and imaginary part of the frequency averaged product of $E_\mathrm{x}$ and $E_\mathrm{z}$ with a negative sign for $V$.

The polarization angle (PA) is
\begin{eqnarray}
    \mathrm{PA} &=& \frac{1}{2} \tan^{-1} \left( \frac{U}{Q} \right), \\
\end{eqnarray}
and we provide the PAs in  the range from $-90^\circ$ to $90^\circ$.

The parameter~$I$ represents the total radiation flux, the parameter~$Q$ represents the polarization component along the $z$- and $x$-axes, the parameter~$U$ represents the polarization component rotated by 45$^\circ$ anti-clockwise from the $z$- and $x$-axes, and $V$ represents the component evolving the orientation of the polarization.
If the radiation is linearly polarized and the polarization plane does not change in time, the linear polarization part $L = \sqrt{Q^2 + U^2}$ is equal to the total radiation flux $L = \eta I$ times the degree of polarization, $\eta$.
If the polarization plane of the linearly polarized waves evolves in time, or changes with distance, then the Stokes parameter $V$ becomes nonzero.\footnote{A plasma containing only linearly polarized waves, but with varying angles and phases will resemble an emitter of partly circularly polarized waves as the sum of the field amplitudes will result in a nonzero Stokes $V$ .} 
In addition, the polarization degree, $\eta$, decreases.

In the $x-z$ plane of Fig.~\ref{fig5}, the Stokes $I$ is high in a circular structure centered around the maximum plasma density ($x \approx -200$, $z\approx0$, denoted by a black plus sign) and
a slightly larger half-moon-shaped region that is mainly located in the bottom part with $x>0$, and has a lower intensity in all Stokes parameters.
Our analysis showed that the circular structures in Stokes $I$ are correlated with the electromagnetic flux of the bunches while the half-moon shape is correlated with electromagnetic wave propagation in the low-density region.

\subsubsection{Polarization properties of escaping waves}
The maps of Stokes $Q$ and $U$ indicate that the orientation of polarization is approximately directed toward the maximum plasma density (black plus sign), that is, along the density gradient.
The polarization orientation in the half-moon shape can be explained as being caused by polarization-sensitive reflection of electromagnetic waves propagating in the low-density Poynting-flux channel.
The waves are reflected off the density gradients as they propagate away from the star in such a way that only the normal component of the electric field is reflected while the tangential component can either propagate through the plasma or is absorbed.
The reflections are significant only for O-mode electromagnetic waves with frequencies below the plasma frequency of the plasma surrounding the low-density channel.
Such wave ducting has been proposed by \cite{Luo2008} and is also known in many other circumstances, for instance, for light waves that are transported in an optical fiber.
The X-mode waves can propagate through the plasma without reflections \citep{Beskin2012}.
The relation between the O- and X-mode waves and the Stokes parameters in both directions is however not straightforward in our simulated scenario.
In addition, the O- and X-mode waves are oblique to the magnetic field, with a small angle between the wave vector $\boldsymbol{k}$ and magnetic field vector $\boldsymbol{B}_\mathrm{dip}$.

The circularly polarized modes $R$ and $L$ do not exist in the plasma because we use the gyro-motion approach for particle tracking, and the computational time step does not resolve the cyclotron motion.
However, the electric field orientation of the linearly polarized waves changes in time together with the changes in density gradients when they pass through the detection plane, resulting in a small Stokes $V$ component.
These changes occur because the O-mode  polarization plane is  determined locally by the density gradients and not by the large-scale magnetic field structure, whose curvature radius is much larger than the typical density variation length.
We discuss this effect in more detail in Sect.~\ref{sec:discuss}.

\subsubsection{Poynting flux properties}
As the time-averaged Stokes parameters do not provide information about the variability of the radio power, we calculate and display in Fig.~\ref{fig6} the polar cap distribution of the mean Poynting flux and its standard deviation.
The B-aligned Poynting flux escaping through the detection plane  shown in Fig.~\ref{fig6}
is overlaid by black lines that indicate the local polarization orientation of the escaping transversal waves.
The magenta circle describes the locations of the last closed field lines on the detection plane.

Analyzing the average Poynting flux and its standard deviation, we found that
the average Poynting flux map is similar to that of the Stokes parameter~$I$, as might be expected.
But the high values of the standard deviation indicate that the flux can temporarily acquire positive as well as negative values. Radiation sometimes propagates toward the stellar surface and at other times it propagates out into the larger magnetosphere.
The standard deviation of the Poynting flux has the highest values in the region around $x \approx -350$\,m, $z = -200$\,m to 200\,m, where the plasma bunches propagate. This region also has a high average Poynting flux, indicating significant temporal evolution in the studied time interval, which we find to be associated with individual outflowing bunches.

\subsection{Stokes parameter profiles across the polar cap}

\begin{figure*}[t!]
    \centering
    \includegraphics[width=\textwidth]{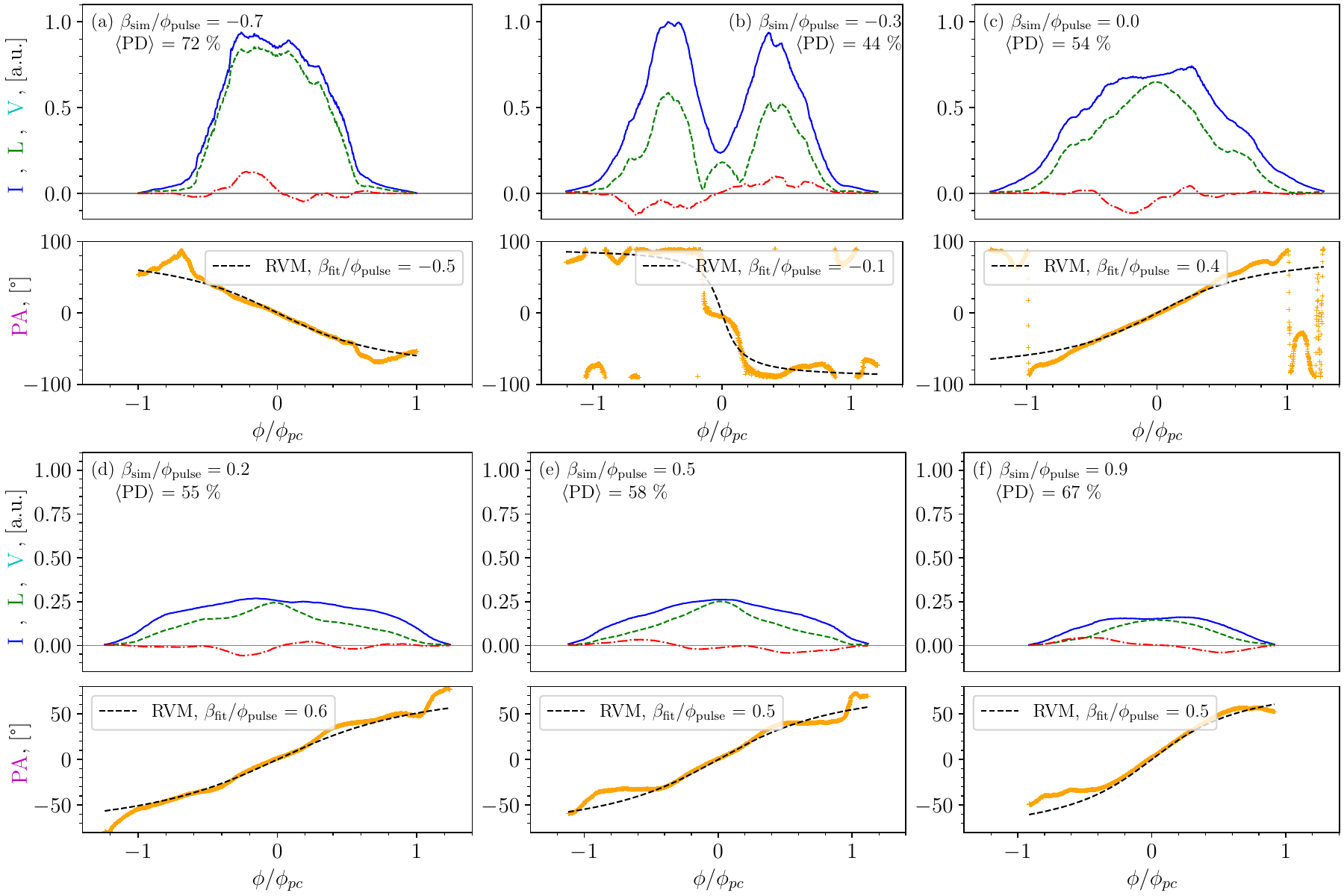}
    \caption{Stokes parameters of the electromagnetic fluxes and their polarization properties as captured in a plane at a height $H$ and averaged over a time interval $\delta t$.
        The trajectory approaches the dipole axis at a minimum angular distance $\beta_\mathrm{sim}$.
        Blue line: Stokes parameter~$I$ of the total flux.
        Dashed green line: Stokes parameter~$|L|$ of the linearly polarized flux.
        Dash-dotted red line: Stokes parameter~$V$ of the circularly polarized flux.
        Orange crosses: PA.
        Dashed black line: The RVM fit (Eq.~\ref{eq:RVM}) of the PA.
        The positions of the profiles (a--f) are shown in Fig.~\ref{fig6} as orange lines.
        The PA scales between plots.
        }     
        \label{fig7}
\end{figure*}

At any instant of time, the radio waves that are received do not come from the entire polar cap; instead, the line of sight of the observer crosses certain parts of the polar cap and samples the radiation and its polarization properties coming from there.
The observer trajectory can be characterized by its latitude $\xi$ in the NS reference frame.

Because the simulation domain covers only the polar cap region and does not include the full radiative transfer through the pulsar magnetosphere, we can estimate the radiation properties only as if they were detected on crossing the polar cap at a trajectory of height $H$ and constant $\xi$, which can be described in terms of a minimum angular distance from the dipole axis, $\beta_\mathrm{sim} = \xi - \iota$.
We selected six trajectories (a)--(f) with $\beta_\mathrm{sim} = -1.5^\circ, -0.8^\circ, 0^\circ, 0.6^\circ, 1.1^\circ$, and $1.7^\circ$ that were 
parallel to the stellar surface.
As the star rotates, they describe trajectories on the $x-z$ plane (orange lines in Fig.~6) that are 
almost straight lines in the polar cap reference frame.
The angular distances, $\beta_\mathrm{sim}$, are selected to cross regions of Poynting flux maxima as well as transitions between these maxima and the polar cap boundaries, potentially representing pulsar observers with different impact angles.

Figure~\ref{fig7} shows the Stokes parameters $|I|$, $|L| = \sqrt{Q^2 + U^2}$, $V$, and the PA (magenta plus signs).
Almost everywhere the plus signs overlap and form thick magenta lines.
The PA is obtained as an angle measured from the
normal vector to the observer's trajectory.
Our simulation time, $T$, is shorter than the typical observer crossing time over the polar cap, $T \ll T_\mathrm{cross} \approx 3$\,ms, and so we cannot
provide information about  variability on longer timescales.
Flux variations on timescales $>T$, which are known from observations, can only be produced by additional magnetospheric plasma or propagation effects that are not included in our model.

There are one or two peaks in the intensity profiles of the Stokes parameter~$I$.
The polarization degree (PD) is estimated as the trajectory-averaged ratio of the polarized fraction to the total flux $I$,
\begin{equation}
    \langle PD \rangle_\phi = \left\langle \frac{\sqrt{Q^2 + U^2 + V^2}}{I} \right\rangle_\mathrm{\phi},
\end{equation}
where $\langle \ldots \rangle_\phi$ denotes the average along the polar cap longitude, $\phi$.
We find large values, exceeding $44\,\%$ for the longitude-averaged PD in all cases and having $\gtrsim75$\,\% at $\phi = 0$.
In all cases, the PA curve has a swing centered at $\phi \approx 0$.

\subsection{Deviations from the rotating-vector model in the polar cap}
The rotating-vector model (RVM) is frequently used for fitting observations \citep{Radhakrishnan1969,Johnston2019}, with the  PA given by   
\begin{equation} \label{eq:RVM}
    \mathrm{PA} = \mathrm{arctan} \left( \frac{\sin \iota \sin (\phi - \phi_0) }{\cos \iota \sin \xi - \sin \iota \cos \xi \cos(\phi - \phi_0)} 
        \right),
\end{equation}
where $\phi_0$ is the longitude of the PA swing between the positive and negative PA and $\iota$   is the dipole inclination angle which is $60^\circ$ in our simulations.
Our fits (in black dashed lines) for $\phi_0,\xi_\mathrm{fit}$ show a good agreement for $\phi_0 = 0$, and the only remaining free parameters is $\beta_\mathrm{fit}=\xi_\mathrm{fit}-\iota$. 
In cases (a) and (c)--(f), the PA follows the RVM model in the central parts of the peak. The RVM function is, however, much smoother than the simulated PA curve which exhibits greater complexity, in which the PA swing also
resembles the observational situation of many pulsars.
At the edges, the signal-to-noise ratio of the simulation data is too low for any conclusions.
Case (b) also shows an additional disturbance in the central part of the peak. 
We find that the differences between the minimum angular distance from the dipole axis obtained from the RVM model $\beta_\mathrm{fit}$ and the observer's minimum angular distance from the dipole axis $\beta_\mathrm{sim}$ are of the order of $-0.4\phi_\mathrm{pulse}$ to $0.4\phi_\mathrm{pulse}$, where $\phi_\mathrm{pulse}$ is the observed radio-pulse width.

After comparing the magnetospheric currents in Fig.~\ref{fig2} and the polarization orientation in Fig.~\ref{fig6}, we find that the time-averaged PA follows the plasma density gradients in the emission region.
Our result indicates that the main assumption of the RVM, namely that the PA is oriented toward the dipole center, is not valid as the PA is instead pointing on the field lines with high plasma densities.
The implication is that, although observed PA swings can be fit by Eq.~\ref{eq:RVM}, the RVM fit cannot reproduce the emission geometry of the neutron star's magnetic field.
The key observational difference would be an offset between the true and the inferred dipole inclination angle which
can be up to a half of the polar cap width.

Stokes V can become nonzero if the orientation of wave polarization is not constant in time but, for example, has a stochastic component. 
Our simulations show that the stochastic component can originate from random density fluctuations of probabilistic pair creation and the subsequent bunching.
Our estimates are associated with a relatively high uncertainty for the Stokes~V parameter, reaching 50\,\% of the Stokes~V maxima.
If we increase the number of data time steps used for the Stokes~V calculation, then the profiles converge to the ones shown in in Fig.~\ref{fig7}.

We do not detect the orthogonal jumps in the PA profiles, which are often observed and thought to be associated with orthogonally polarized plasma-wave modes \citep{Smits2006}.
It is possible that such features result from subsequent propagation effects in the magnetosphere that are not included by our simulations; we discuss this point further in Sect.~\ref{sec:propagation}.

\subsection{Spectra of Stokes $I$ and polarization degree}

\begin{figure}[t!]
    \centering
    \includegraphics[width=0.45\textwidth]{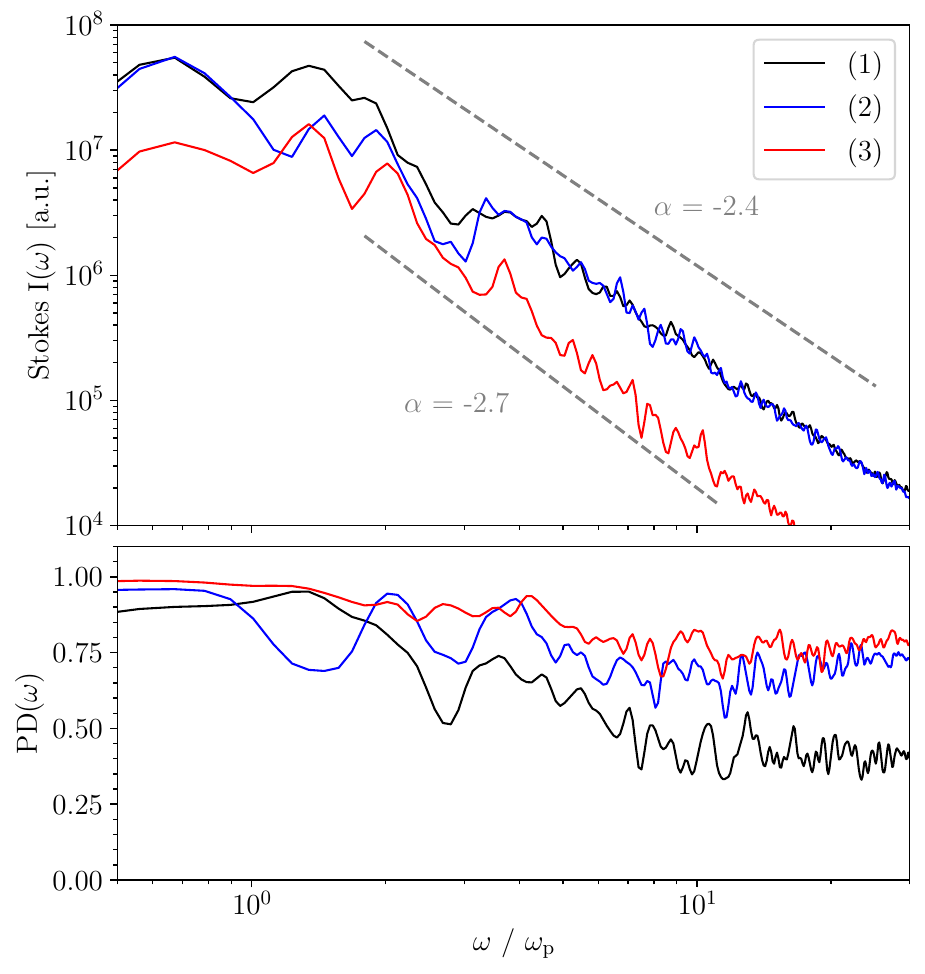}
    \caption{Top panel: Power spectra of Stokes $I$ parameters (right) for three positions in the polar cap (left) with the frequency scaled to the plasma frequency $\omega_p=1.6\cdot10^9\ \mathrm{rad\, s^{-1}}$.
        The analyzed regions are denoted as black boxes in Fig.~\ref{fig5}a. Bottom panel: Polarization degree as a function of frequency at the same locations as above. 
        }     
        \label{fig9}
\end{figure}

We analyze the spectra in Fig.~\ref{fig9} for three selected regions in the polar cap, which are indicated in Fig.~\ref{fig5}a.
Two of the regions are selected because they contain the radiation associated with plasma bunches and one region lies in the Poynting-flux channel.
All three spectra have a broadband character and can be characterized by maxima around $\omega/\omega_\mathrm{p} \approx 1$ and power-law spectra for $\omega/\omega_\mathrm{p} > 1$, where $\omega_\mathrm{p} \equiv \omega_\mathrm{p}(n_\mathrm{GJ,axis})$.
The power-law indices are $-2.4 \pm 0.2$ for the radiation associated with the plasma bunches and $-2.7 \pm 0.2$ for the radiation in the Poynting flux channel.
The different spectral indices may indicate that the emission processes and propagation in the plasma bunches could be different from those in the Poynting flux channels. 
In addition, these indices are lower than those obtained in the 2D simulations by \citet{Benacek2024b}.
We explain the difference with the better resolution and the higher maximum particle density in the 2D simulations 
and the fact that the 2D case allows the generation of more intensive waves at higher frequencies.

Our simulations do not indicate any radius-to-frequency dependence,  as 
the escaping power at all frequencies is generated instantaneously by the plasma bunches or the oscillating electric gap in the polar cap region.

\subsection{Escaping electromagnetic power}

\begin{figure}[t]
    \centering
    \includegraphics[width=0.45\textwidth]{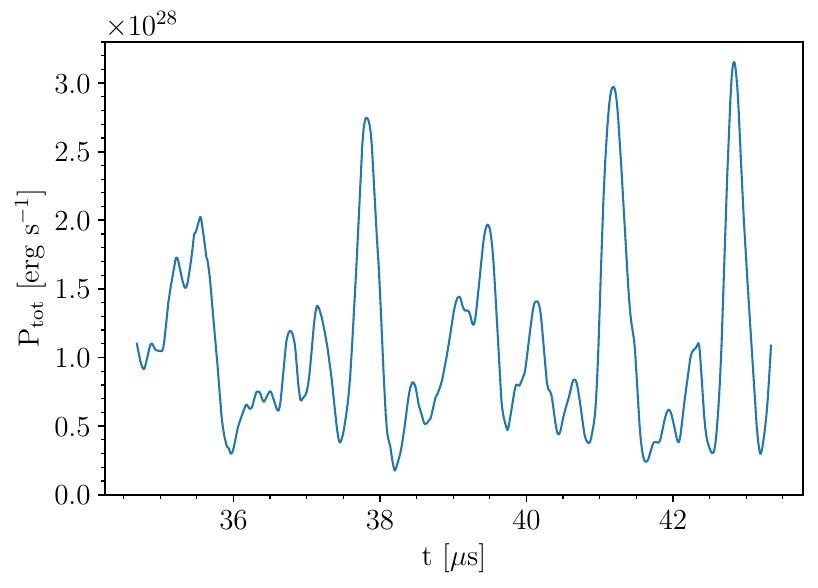}
    \caption{Time evolution of total electromagnetic power estimated from the Poynting flux escaping from the polar cap.
        }     
        \label{fig8}
\end{figure}

The emitted radio power is estimated by integration of the escaping Poynting flux over the virtual plane. 
The power is then scaled from the reduced simulation quantity $S'$ to the observer value $S$ using $S=\zeta^2 S'$, as outlined in  Sect. ~\ref{ssect:scaling} and \cite{Benacek2021b}, yielding $1.1\times10^{28}$\,erg\,s$^{-1}$. Note that a part of the emitted radio power may be absorbed by the plasma during propagation through the magnetosphere.
The total predicted power is in the range of the measured powers of pulsars, $10^{25}$--$10^{29}$\,erg\,s$^{-1}$ \citep{lorimer_handbook_2005,Philippov2022}
and hence just a small fraction of the total pulsar spin-down power is visible in the radio output. 
The spin-down power of our simulated pulsar can be estimated as \citep[Chapter~6]{Condon2016}
\begin{equation}
    L \approx \frac{B_\mathrm{dip,axis}^2 R_\star^6 \Omega^4 \sin^2 \iota}{6 c^3} \approx 1.8\times 10^{33}\,\mathrm{erg}\,\mathrm{s}^{-1},
\end{equation}

Figure~\ref{fig8} shows the time evolution of the rescaled power of the escaping electromagnetic radiation.
The electromagnetic emission produced by the plasma bunches displays strong short-timescale intensity fluctuations.
This profile has large spikes separated in time by $1-2\,\mu$s (about $ 300-600$\,m in space), corresponding to the typical time intervals and distances between consecutive large plasma bunches created in the electric gap region.
This offers a natural explanation for microsecond variability and variable nonisotropic emission from the polar cap.

\section{Discussion} \label{sec:discuss}
In this study, we use PIC simulations to investigate the polarization properties of electromagnetic waves emitted by pair cascades in pulsar polar caps.
As the simulations self-consistently describe the particle-wave and wave-wave interactions, we can obtain the properties of the emitted radio waves from the generated coherent electromagnetic fields.
To our knowledge, this is the first time 3D PIC simulations have been used to obtain the polarized radio emission described by the Stokes parameters $I, Q, U,$ and $V$.
The similarity of polarization properties of the escaping radiation to pulsars with high spin-down power (defined as $R_1$-category in \citet{Mitra2016b} as pulsars with high linear polarization and RVM-like swing) suggests that the subsequent magnetospheric propagation may only have a weak impact on the observed radio properties.

The ability to model the radio power and spectrum, the pulse profiles, polarization curves, and high temporal variability is a strong point of our model.
More complex profiles on microseconds scales than the model provides can be attributed to a nondipole structure of the magnetic fields in the polar cap, complicating the magnetospheric current profile, $j_\mathrm{mag}/j_\mathrm{GJ,axis}$, and to radiative-transfer effects.
The changes on longer timescales might be attributed to large-scale variability in the magnetospheric structure.

Our current model cannot address how the Poynting flux continues to propagate through the magnetosphere, and to what extent flux will be absorbed or reflected later on.

\subsection{Coherent electromagnetic waves escaping from the polar cap }
Our 3D simulations confirm the findings of 2D simulations that the escaping electromagnetic waves represented by the Poynting flux are related to the dense plasma bunches \citep{Philippov2020,Cruz2021b}.
The absorption of the Poynting flux associated with plasma bunches turns out to be lower in the 3D simulations than in the 2-D  case of \citet{Benacek2024b}.
This may be the result of the lower grid resolution or the difference in the dimensionality, but we do not know which is the more decisive factor at this stage.

The waves that escape from the dense plasma bunches can then propagate without significant absorption in Poynting-flux channels that are located in regions of low plasma density and open magnetic field lines \citep{Benacek2024b}.
Such a Poynting flux channel is defined by the relativistically modified plasma frequency being below the frequency of the electromagnetic waves. The magnetospheric currents are small and do not allow significant pair production, typically $1 > j_\mathrm{mag}/j_\mathrm{GJ} \geq 0$.

The escaping radiation forms single- or double-component profiles when averaged over 8.7\,$\mu$s, depending on the observed latitude across the polar cap.
The polar cap can produce diverse types of observed pulsar profiles under the assumption that the radiation is not too strongly affected by its propagation through the magnetosphere.
The distance from the magnetic axis is derived from the swing of the PA profile, but the estimates are not consistent with the RVM, as we discuss below.

The Poynting flux escaping the polar cap fluctuates according to the quasi-periodicity of the pair discharge.
These oscillations may be interpreted as microstructures that are observed in radio pulse profiles \citep{Cordes1979}.
It is not straightforward to compare the timescales between observations and our model. In our model, the observed microstructure timescale is reproduced as being as short as a few microseconds \citep{Tian2025}.
High temporal variability of the radio flux, similar to our model results, is indeed typical for single pulses \citep{Cao2024b}. 

We do not detect any orthogonal polarization modes and jumps in the emission in our simulations, suggesting that there is only one dominant polarization mode of electromagnetic waves escaping the polar cap.
These modes probably appear during the subsequent  radiative transfer as a result of wave mode conversion and the scattering difference between the X- and O-mode polarized waves \citep{Karastergiou2002,Melikidze2014}.

\subsection{Polarization properties}

Localized dense plasma bunches result from the pair creation process (see Fig. \ref{fig3} ). Their nonuniform charge distribution creates strong electric fields (Fig. \ref{fig4}, left panel) that are preferentially aligned with the bunch density gradients as was already demonstrated by \cite{Philippov2020}. These fields give rise to large-amplitude waves in the $\boldsymbol{k}-\boldsymbol{B}$ plane, and the resulting electromagnetic waves are therefore classified as oblique O-mode waves.
Radio waves escape from the dense pair-plasma clouds into the voids of the turbulent flow with subsequently high Poynting fluxes there (Fig. \ref{fig4}, right panel). These escaping waves are strongly linearly polarized and 
we find that their PA orientation is not primarily determined by the direction of the magnetic field as assumed for the RVM,  but that it is preferentially aligned with the density gradient of the plasma flow.
\subsubsection{The characteristic frequency of the emitted radio spectrum}
We find that the energy distribution of our simulated pair plasma particles is very wide. Following \cite{Rafat2019a}, this allows us to estimate the energy distribution and volume averaged Lorentz factor from the simulation data as  $\langle \gamma^{-3} \rangle \simeq \langle \gamma \rangle^{-1} \simeq \left(3.6\times 10^{7}\right)^{-1}$, a value between the Lorentz factor of primary and secondary particles.
All regions of the polar cap contribute to the observable power spectrum, and the average will therefore include a wide range of densities and Lorentz factors.
It is however dominated by particles with very high Lorentz factors even if their number densities are small.
The multiplicity $\kappa$ also varies through the simulation volume, but the radio emission is created mainly in the high $\kappa$ regions.
Together with a multiplicity $\kappa\simeq 10^5$ \citep{Timokhin2019}, we obtain 
the typical plasma frequency in the simulation volume can be given as
\begin{equation}
    \omega_\mathrm{p}=\sqrt{\frac{4\pi \kappa n_\mathrm{GJ,axis}}{\langle \gamma \rangle m_{e}}}\ \cong 1.6\times 10^{9}\,\mathrm{rad\,s}^{-1} \cong 250\,\mathrm{MHz}.
\end{equation}
The emitted radio  spectrum is approximately flat up to a break at roughly 250~MHz and then continues as a softer power-law, similar to many observed pulsar spectra. The specific locations of the observed spectral breaks depend on the characteristics of the pair plasma of individual pulsars and can also be influenced by subsequent free-free absorption in the interstellar medium.

\subsubsection{Production and interpretation of nonzero Stokes $V$}
The polarization orientation has a stochastic variation resulting from random fluctuations in
the pair creation process and the consequent bunching effects. It is therefore changing slightly along  the plasma bunch as the bunch propagates through the detection plane.  This causes the time-averaged Stokes parameter $V$ to acquire a nonzero residual value and introduces a decrease of the polarization degree.

\citet{Radhakrishnan1990} proposed that 
an antisymmetric swing of Stokes~V across the pulse profile can be a signature of an intrinsic property of the emission mechanism.
Figure~\ref{fig7}c also shows a symmetric profile, indicating that both types of Stokes~$V$ profiles could originate in the polar cap. The observed Stokes $V$ could, however, be interpreted as a signature of how much the imprinted plasma structures vary in the emission source.

Furthermore, some pulsars manifest only one preferential handedness of the circular polarization in the pulse \citep{Johnston2019}.
Though our model produce regions with a preferential polarization handedness in the polar cap, the handedness physical origin remains an open question.

\subsubsection{Radiation depolarization effects}
The decreasing polarization degree with increasing frequency can be explained by the high amplitude of plasma charge fluctuations at smaller scales.
They cause the PA to oscillate faster and more frequently than at longer wavelengths, leading to a lower polarization degree and a relative increase of the nonzero Stokes~$V$ component.
Additional depolarization effect can occur during the radiative transfer \citep{vonHoensbroech1998}.

\subsubsection{PA orientation of O-mode waves}
Plasma density fluctuations are the main factor for determining the orientation of PA, and they are therefore different from  the classical assumption that the O-mode PA must align with the magnetic field.
This can be understood by realizing that the typical scales of plasma fluctuations and the generated electromagnetic waves differ greatly from the typical scales of environmental changes.
The waves originate from charge-density variations at wavelengths $\lambda$, which are not aligned with the magnetic field. The magnetic field variation on the wavelength scale is negligible and, in fact, can be considered homogeneous for much larger regions, because $\lambda \ll \rho$.
In addition, the classical theory of wave modes in plasma is not fully applicable as the theory assumes only small, linearized perturbations of charge density and related quantities and homogeneous magnetic fields, and these assumptions are not valid in the pulsar polar cap.

The radiation escaping from the oscillating electric gap is of mixed polarization, but mostly composed of O-modes at low heights where it escapes through the Poynting flux channel.
When the waves propagate along the channel, they eventually reach the dense plasma wall of the channel where the plasma frequency exceeds the wave frequency.
The O-modes are then reflected back into the channel which acts as a waveguide \citep{Luo2008} as the result of the increasing refraction index, while X-mode waves, having only small amplitudes in our case, continue to propagate without reflection.

To propagate a significant distance in the channel, the waves must have been reflected at the channel boundaries, resulting in O-mode polarization of the escaping emission. But waves with higher frequencies suffer fewer reflections, as the plasma density decreases with height, and will consequently be less polarized (Fig. \ref{fig9} lower panel). This decrease of polarization with increasing frequency is well known from observations \citep{ Xilouris1996,Liu2019}.

\subsection{Reappraisal of RVM estimates }
The RVM is often used for the estimation of the angle between the magnetic and rotation axes and the observer's line of sight, based on the assumptions that the magnetic field is dipolar and that the linear polarization angle is dictated by the magnetic field line structure.

Although our modeled PA swings can formally be fit quite well with the RVM (Eq. \ref{eq:RVM} ) for most viewing angles, and especially for those close to the magnetic axis, the angular distances obtained from the fit, $\beta_\mathrm{fit}$, are not the same as the trajectory latitudes, $\beta_\mathrm{sim}$, in the polar cap.
Assuming that the radio pulse had an observed width, $\phi_\mathrm{pulse}$, the difference between the fit and obtained angles can as high as $\pm0.4 \phi_\mathrm{pulse}$.
For an estimated observed angular pulse width of a few tens of degrees, the error in the estimation of the angular distance from the dipole for a pulsar with $\iota = 60^\circ$ can therefore be around ten degrees.
This may potentially lead to systematic errors when estimating the inclination angles for pulsars from observations and casts doubts on the validity of the inclination angles of pulsars estimated from their observed PA.
There may not be a good match for the PA swing in the RVM fit for radiation originating in outer regions of the polar cap because the radiation can also interact with the plasma at closed field lines. 
Moreover, the correction might be also applicable to pulsars with interpulses because the structure of the emission beam from poles is similar, under the assumption of a dipolar magnetic field and that the whole polar region is located above one hemisphere.

It is nevertheless still possible to use the observed PA curves for estimates of the magnetic inclinations and viewing angles by using the direction of plasma-density gradients derived from global magnetospheric current modeling like that by \cite{Gralla2017} that we have used for our simulation. That will undoubtedly require more computational efforts, but given the fact that our simulated PA profiles show greater complexities than the simple RVM swings, it also provides us with a chance of better fits for many of complex PA profiles that have been observed.

\subsection{Effects of magnetospheric propagation} \label{sec:propagation}
Our results show that the radiation escaping the polar cap is strongly linearly polarized with a high polarization degree and that the radiation PA follows the density gradient already when escaping in the polar cap.
Any detected strong circular polarization can be attributed to the radiative transfer rather than to its direct formation in the polar cap \citep{Hakobyan2017} as the gap emission itself has only a very low residual circular polarization.
Moreover, the radiation within the polar cap cannot be represented by an initially unpolarized uniform ``slab''  across the polar cap, as was considered in some radiative transfer models \citep{Petrova2009}.
Instead, the intensity structure follows a combination of a half-moon shape of the Poynting-flux channel and a circular shape of the radiation associated with the relativistically outflowing plasma bunches.

\subsection{Comparability with observations }
One of the aspects that could be compared with observations is the statistical distribution of RVMs \citep{Johnston2023} that can be fit to PA swing in simulations with a statistical sample of inclination angles.
However, this would require providing numerous simulations, making it a hard computational task.
In addition, multipolar components of the stellar magnetic field in the polar cap can further modify the PA profiles and their possible fits by the RVM.
Furthermore, we expect that the radiation of pulsars with a high circularly polarized fraction is strongly processed during the radiative transfer; their PA profiles therefore do not follow the RVM.

Most of the escaping power is generated as broadband radiation by the plasma bunches or the oscillating electric gap in the polar cap, and waves at all frequencies simultaneously propagate toward the observer. 
A radius-to-frequency relation of their origin in the polar cap is ruled out by our model, and this is consistent with the observations \citep{Hassall2012,Hassall2013}.
Counter examples to the predictions of radius-to-frequency mapping  were also given by \citet{Posselt2021} who showed that a subset of the pulsar population has pulse widths that widen with increasing frequency.

The method of our simulations allows the direct prediction of all four Stokes parameters of the polar cap radio emission for any chosen pulsar and any combination of magnetic inclination and viewing angles, and thus can be used to test the emission model.

\section{Summary and conclusions} \label{sec:conclude}
Our PIC simulations are the first full 3D simulations of the polar cap pair-production region that yield comprehensive and detailed information  about the major features of pulsar radio emission that is largely consistent with observational evidence. Our model can explain the following characteristics of pulsar radio emission:
\begin{enumerate}
    \item The radiation is mainly linearly polarized, with the polarization angle showing an S-shaped swing.
    \item The PA can be fit by the RVM model, but the resultant angles are inaccurate as the orientation of the polarization vector is given by the direction of the plasma density gradient and not by the center of the magnetic dipole.
    \item The degree of linear polarization decreases with frequency.
    \item There is also weak circular polarization in the escaping radiation.
    \item There is no radius dependence. The waves at all frequencies are generated instantaneously in a small region.  
    \item The radio flux escapes along channels of low plasma density that act as waveguides.
    \item The total emitted radio power amounts to about $10^{28}\ \mathrm{erg ~ s^{-1}}$ (Sect. 3.6) for a spin-down power $1.8\times 10^{33}\,\mathrm{erg}\,\mathrm{s}^{-1}$.
    \item The radio spectrum follows a soft power law above a break at about 250~MHz (Sect. 3.5).
    \item The radio emission fluctuates on microsecond timescales (Sect. 3.6). 
    \item The pulse-profile morphology is similar to the observed pulse shapes and depends on the viewing angle.
\end{enumerate}

The polarization properties of the coherent electromagnetic radiation emerging from the polar caps of pulsars are crucial for calculating the subsequent radiative transfer through the magnetosphere and for obtaining specific predictions on observable radio properties. 
In agreement with the 2D simulations by \citet{Benacek2024b}, we found that the radiation is associated with plasma bunches that are the result of nonstationary pair production and that the radiation can propagate freely along Poynting-flux channels of low plasma density.
The properties of the radiation from the modeled polar cap are similar  to those observed of energetic pulsars with a strong linear polarization and a single PA swing that can be fit by the  RVM  \citep{Mitra2016b}. 
The similarity of our simulation results to the observed characteristics suggests that the escaping radiation of these pulsars is not strongly reprocessed during its passage through the magnetosphere.
Many of the differences between our model and observations of other pulsars can be attributed to wave-mode conversion and wave scattering during radiative transfer.

In our model, the radiation generated in the polar cap is broadband with no radius-to-frequency mapping. Any profile evolution with frequency must therefore be a propagation effect.

The radiation flux and PA profiles depend on the position in the polar cap.
The radio flux can form one and two peaks and is strongly polarized, with a typical total polarization degree of over 44\,\% for all studied positions, and it exceeds 75\,\% in the pulse center.
We propose that the range of other polarization features in observations \citep{Posselt2023,Oswald2023b} might be explained by subsequent propagation effects that were not part of the simulation.

Although the PA of the radiation in the polar cap can be fit by the RVM profile, the latitude angles obtained from the RVM fit, $\beta_\mathrm{fit}$, do not correspond to the real angular distance of the observer, $\beta_\mathrm{sim}$, from the magnetic axis.
The angular difference can reach half of the detected pulse width.
The radiation PA in the polar cap is directly related to the magnetospheric currents that are set as boundary conditions  in the simulation, and it follows the plasma density gradients in the polar cap.
Assuming that the PA is not significantly changed during propagation through the magnetosphere, we can determine the plasma density and magnetospheric current profiles along the observed path by  mapping the observed PA profiles.
This might allow us to test the current profiles obtained from magnetospheric neutron star models by comparing them with those derived from the observations of the radio flux and PA.

Although there is initially no circularly polarized mode in the polar cap plasma because of the extremely strong magnetic fields, we report that the calculated Stokes parameter~$V$ of the escaping radiation is nonzero.
This a feature of the method of Stokes-$V$ calculations, which only consider the change in the orientation of electric oscillations with time, without regard to the cause of a change in the orientation.
The observed flux in the Stokes-$V$ parameter, often denoted as circularly polarized intensity in observations, is traditionally related to a circularly polarized wave mode shaped by cyclotron motion, but it can just as well be produced by quick, stochastic changes in the O-mode wave orientation and the associated gradients in the plasma density.

For detailed quantitative analyses of the radiative properties, we require simulations with higher resolutions than in our case.
It would also be of interest to model the propagation of our simulated radio emission in a large-scale magnetosphere that matches the current profiles given as boundary conditions for our simulations, and to compare those to observations. PSR J1906+0746 is known to show relativistic spin precession, and its polarization observations at varying viewing angles \citep{desvignes_radio_2019} might be used to test our model of the polar cap radio-emission processes combined with the propagation in a full-scale magnetosphere model.

\begin{acknowledgements}
We are grateful for the suggestions by an anonymous referee which helped us to improve our manuscript. We acknowledge the developers of the ACRONYM code (Verein zur F\"orderung kinetischer Plasmasimulationen e.V.).
We also acknowledge the support by the German Science Foundation (DFG) project BE~7886/2-1. 
This project has received funding from the European Union’s Horizon Europe
research and innovation programme under the Marie Skłodowska-Curie grant
agreement No 101203963.
LSO acknowledges support from the EPSRC Stephen Hawking Fellowship grant EP/Z534730/1.
The authors gratefully acknowledge the Gauss Centre for Supercomputing e.V. (\url{www.gauss-centre.eu}) for partially funding this project by providing computing time on the GCS Supercomputer SuperMUC-NG at Leibniz Supercomputing Centre (www.lrz.de), projects pn73ne and pn52ku.
\end{acknowledgements}

\bibliographystyle{aa}
\bibliography{references}

\end{document}